\documentclass[pre,aps,twocolumn]{revtex4}
\usepackage{epsfig}
\usepackage{bm}
\newcommand{\B}[1]{{\bm{#1}}}

\usepackage[latin1]{inputenc}
\newcommand{\beq}{\begin{equation}}
\newcommand{\eeq}{\end{equation}}
\newcommand{\bea}{\begin{eqnarray}}
\newcommand{\eea}{\end{eqnarray}}
\begin{document}
\title{Statistical Physics of Fracture Surfaces Morphology}
\author{Eran Bouchbinder, Itamar Procaccia and Shani Sela}

\affiliation{Dept. of Chemical Physics, The Weizmann Institute of
Science, Rehovot 76100, Israel}
\begin{abstract}
Experiments on fracture surface morphologies offer increasing
amounts of data that can be analyzed using methods of statistical
physics. One finds scaling exponents associated with correlation and
structure functions, indicating a rich phenomenology of anomalous
scaling. We argue that traditional models of fracture fail to
reproduce this rich phenomenology and new ideas and concepts are
called for. We present some recent models that introduce the effects
of deviations from homogeneous linear elasticity theory on the
morphology of fracture surfaces, succeeding to reproduce the
multiscaling phenomenology at least in 1+1 dimensions. For surfaces
in 2+1 dimensions we introduce novel methods of analysis based on
projecting the data on the irreducible representations of the SO(2)
symmetry group. It appears that this approach organizes effectively
the rich scaling properties. We end up with the proposition of new
experiments in which the rotational symmetry is not broken, such
that the scaling properties should be particularly simple.

\begin{widetext}
{\it It is a privilege to dedicate this paper to Pierre Hohenberg and Jim Langer who contributed,
respectively, decisive ideas to scaling concepts in statistical physics and to the study of fracture.
 It is our hope that the marriage of these two issues in the present paper may give them some degree of pleasure.}
\end{widetext}

\end{abstract}
\maketitle
\section{Introduction}

The failure of materials is a challenging interdisciplinary problem
of both technological and fundamental interests. From the
technological point of view, the understanding of the failure
mechanisms of materials under various external conditions may
improve dramatically the integrity of structures in a wide range of
applications. From the theoretical point of view, the understanding
of the way materials fail entails the development of new
mathematical methodologies and necessitates the introduction of new
concepts in non-linear and solid state physics. The modern
development of the field as a scientific discipline initiated with
the pioneering work of Griffith \cite{20Griff} who identified the
importance of defects in determining the strength of materials.
These defects act as {\em stress concentrators} in the sense that
the typical stress near the defect can be much higher than the
applied stress. In that way the strength of materials is highly
reduced, explaining the long lived conflict between theoretical
strength estimations and experimental observations. In order to
understand the way defects affect the failure of materials we first
introduce a simple scaling argument in the framework of equilibrium
thermodynamics.

\begin{figure}[here]
\centering \epsfig{width=.25\textwidth,file=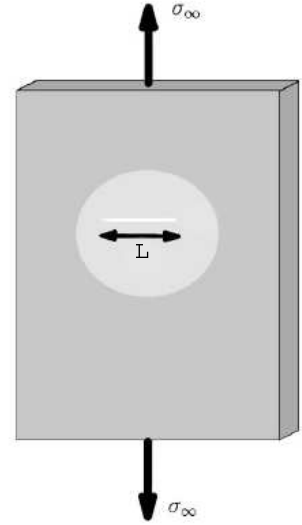} \caption{A
material under the application of a uniform external stress
$\sigma_{\infty}$ at its far edges in the presence of a crack of
length $L$ that is cutting through the sample. The shaded region
represents the typical area in which the potential energy density is
changed relative to the uniform stress state.} \label{Sketch}
\end{figure}

In Fig. \ref{Sketch} we consider a material under the application of
a uniform external stress $\sigma_{\infty}$ at its far edges. In
addition, the material contains a {\em single} defect that is
assumed here to be a crack of length $L$ that is cutting through the
material. A crack is a region whose boundaries cannot support
stress. In the absence of a crack the material is uniformly stressed
with an associated potential energy density ${\cal E}_p$
\begin{equation}
{\cal E}_p \sim \frac{\sigma^2_{\infty}}{E} \ , \label{Up}
\end{equation}
where $E$ is Young's modulus and it is assumed that the material is
linear elastic, resulting in a second-order energy density. The
presence of a crack of length $L$ releases the stresses in an area
of the order of $\sim L^2$ (shaded area in Fig. \ref{Sketch}),
resulting in a reduction $\Delta U_p$ in the potential energy per
unit material width
\begin{equation}
\Delta U_p = -c_p\frac{\sigma^2_{\infty} L^2}{E} \ , \label{reduc}
\end{equation}
where $c_p$ is a dimensionless factor. On the other hand, the
very existence  of the crack is associated with free material surface.
Assuming that the energy cost per unit area of free surfaces is $\Gamma$,
the generation of a crack of length $L$ results in an increase in the
surface energy per unit width by an amount $\Delta U_s$,
\begin{equation}
\Delta U_s =\Gamma L \  . \label{surf}
\end{equation}
The total energy change per unit width is
\begin{equation}
\Delta U =  -c_p\frac{\sigma^2_{\infty} L^2}{E} + \Gamma L\ .
\label{tot}
\end{equation}
This equation tells us that for small values of $L$ the formation of a crack
is costly ($\Delta U > 0$) whereas longer cracks are energetically favorable
 ($\Delta U < 0$).  Actually, once a critical length is achieved the crack tends to
increase indefinitely until the material completely fails. This
non-equilibrium catastrophic crack propagation is at the essence of
the failure of material. The two different regimes described above
are separated by a Griffith critical length
\begin{equation}
L_G \sim   \frac{E \Gamma}{\sigma_{\infty}} \ , \label{Griffith}
\end{equation}
which is shown to be a combination of material properties ($E$ and
$\Gamma$) and external loading conditions ($\sigma_{\infty}$).

The thermodynamic argument predicts that crack propagation should always
be catastrophic once initiated. In fact, although fast crack
propagation is common, there are many situations in which the crack
evolves quasi-statically. The point to stress is that whatever is the mode
of crack growth, the argument
exemplifies the multi-scale nature of the phenomenon; the potential
energy released from the large scales dissipates in a very
localized region near the crack tip where new crack surfaces are
generated.

The technical discussion of crack propagation is classically done in
the context of ``linear elasticity fracture mechanics''. The state
of deformation of the material is described by the displacement
field ${\bf u(r)}$. This field consists of three translational
degrees of freedom, where the local rotational degrees of freedom
are neglected \cite{Cosserat}. To develop a theory that is both
translational and rotational invariant one defines the strain tensor
$\epsilon_{ij}$, which is the symmetric part of the gradient of
${\bf u(r)}$
\begin{equation}
\epsilon_{ij} \equiv \frac{1}{2} \left(\partial_i u_j + \partial_j
u_i \right) \ . \label{strain}
\end{equation}
In a material that is homogeneous and isotropic, only the invariants
of $\epsilon_{ij}$ can appear in the expression for the strain
energy density ${\cal E}$. Restricting the analysis to a linear
theory, i.e. to a quadratic strain energy density, we arrive at
\begin{equation}
{\cal E} = \frac{1}{2} \left(2 \mu~{\rm Tr}(\B\epsilon^2)+ \lambda
~{\rm Tr}^2(\B\epsilon) \right) \ , \label{strainEnergy}
\end{equation}
where the Lam\'e coefficients $\mu$ and $\lambda$ are material
parameters that are related to the more common engineering constants
$E$ (Young's modulus) and $\nu$ (Poisson's ratio). The stress tensor
$\sigma_{ij}$ is related to the strain tensor via
\begin{equation}
\sigma_{ij} = \frac{\partial {\cal E}}{\partial \epsilon_{ij}} \ ,
\label{stress}
\end{equation}
which leads to a linear stress-strain relation
\begin{equation}
\sigma_{ij} = 2 \mu \epsilon_{ij} + \lambda \delta_{ij}
\epsilon_{kk} \ . \label{Hooke}
\end{equation}
The stress field is just the local force per unit area. Therefore,
the Newton's equations of motion for a unit volume of mass density
$\rho$ are given by \cite{86LL}
\begin{equation}
\frac{\partial \sigma_{ij}}{\partial x_j} = \rho \frac{\partial^2
u_i}{\partial t^2} \ . \label{EOM}
\end{equation}
Substituting Eq. (\ref{Hooke}) in the last set of equations we
arrive at the Lam\'e equation
\begin{equation}
(\lambda+\mu)\B\nabla(\B\nabla\cdot {\bf u})+\mu \nabla^2 {\bf u}
= \rho \frac{\partial^2 {\bf u}}{\partial t^2}\ . \label{lame}
\end{equation}

\begin{figure}
\centering \epsfig{width=.48\textwidth,file=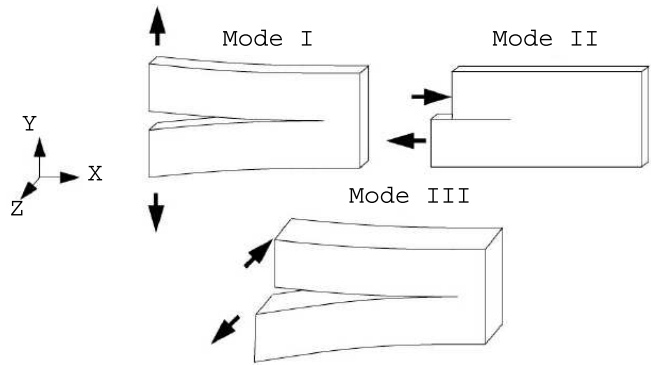}
\caption{The typical symmetry modes of fracture, see text for more
details.} \label{Modes}
\end{figure}

Up to now the field equations for the medium were constructed
disregarding the effects induced by the presence of a crack. At the
level of linear elasticity fracture mechanics, the crack is
introduced as boundary conditions. As was mentioned before, a crack
is a region whose boundaries are free surfaces. Denote the unit
normal to the crack surface at any point by $\hat{{\bf n}}$; the
boundary conditions on the crack surface are
\begin{equation}
\sigma_{ij}n_j = 0 \ . \label{BC}
\end{equation}
These boundary conditions introduce non-linearities into the problem
even if the field equations themselves are linear. This is a
major source of mathematical difficulties.

 It is conventional to
decompose the stress field under general loading conditions to three
symmetry {\em modes} with respect to the fracture plane. These are
illustrated in Fig. \ref{Modes}. In mode I the crack faces are
displaced symmetrically in the normal direction relative to the
fracture $xz$ plane, by tension. In mode II, the crack faces are
displaced anti-symmetrically relative to the fracture $xz$ plane, in
the $x$ direction, by shear. In these two modes of fracture the
deformation is in-plane ($xy$). In mode III, the crack faces are
displaced anti-symmetrically relative to the fracture plane, in the
$z$ direction, by shear. This is an out-of-plane fracture mode.

The dissipation involved in the crack growth, quantified by the
phenomenological material function $\Gamma$, is assumed to be highly
localized near the crack front. Therefore, one is interested in the
near crack front fields. Within linear elasticity theory these are
actually singular. To see this consider first a quasi-static
infinitesimal extension $\delta L$ of the crack. Denoting the stress
field near the tip by $\sigma (r)$,  the energy (per unit width)
released from the linear elastic medium $\delta U$ is
\begin{equation}
\delta U \sim \int_0^{\delta L} \!\frac{\sigma^2(r)}{E}~r~dr \ .
\end{equation}
This amount of energy is invested in creating new crack surfaces
whose energy cost (per unit width) is $\Gamma \delta L$. Therefore,
we must have
\begin{equation}
\sigma (r)\sim \frac{1}{\sqrt{r}} \ . \label{StressScaling}
\end{equation}
The inverse square-root singularity seen in Eq. (\ref{StressScaling}) exists
also in the fully dynamic case. In fact, by asymptotic
expansion of the local stress tensor field near the crack front, it
can be shown that the leading term is given by a sum of three
contributions corresponding to the three modes of fracture
\cite{98Fre}
\begin{eqnarray}
\sigma_{ij}(r,\theta,t) &=& \nonumber \\
K_{_{\rm I}}(t) \frac{\Sigma^{^{\rm I}}_{ij}(\theta,v)}{\sqrt{2\pi
r}}&+&K_{_{\rm II}}(t) \frac{\Sigma^{^{\rm
II}}_{ij}(\theta,v)}{\sqrt{2\pi r}} + K_{_{\rm III}}(t)
\frac{\Sigma^{^{\rm III}}_{ij}(\theta,v)}{\sqrt{2\pi r}} \ , \nonumber \\
\label{sigma}
\end{eqnarray}
where $(r,\theta)$ are local polar coordinates system, $t$ is time,
$v(t)$ is the local instantaneous front velocity, $\B
\Sigma^{(i)}(\theta,v)$ are universal and $K_{(i)}$ are the stress
intensity factors. The stress intensity factors are non-universal
functionals of the loading conditions, sample geometry and crack
history. The predicted singular behavior is, of course, not physical
and there must exist mechanisms to cut off this apparent
singularity. Nevertheless, there are many physical situations for
which the size of the region where linear elasticity breaks down is
small compared to other relevant lengths. Therefore, the stress
intensity factors are very important physical quantities and the
singularity may be retained in many models. This singularity is a
source of mathematical difficulties and physical riddles. The stress
concentration quantified by the stress intensity factors shows that
the material near a crack front experiences extreme conditions; the
response to these conditions is far from being well-understood.

Up to now we have considered the case of a single straight crack in
an otherwise linear homogeneous medium. A more realistic description
of materials includes other sources of heterogeneity which exist in
every material at some level. The effect of distributed sources of
``disorder'' calls for a statistical treatment. In developing such
an approach one should understand the interplay between various
fields and the material disorder, resulting in a complex
spatio-temporal behavior. This complexity was studied extensively in
the last decade, accumulating a wealth of relatively high precision
experimental data and offering a real challenge for the theoretical
physicist \cite{99FM}. In this paper we focus on one important
aspect of the problem: the statistical physics of the morphology of
quasi-static fracture surfaces. This subject has attracted a lot of
interest recently and became a very active field of research
\cite{03Bou}. The pioneering experimental work described in
Ref.~\cite{84MPP} drew attention to the fact that fracture surfaces
are rough graphs in 2+1 (1+1) dimensions when the broken sample is
three dimensional (two dimensional) and therefore might have
anisotropic scaling properties. Suppose that the surface is
described by its height $h(x,y)$ at position $(x,y)$ in a smooth
reference plane. Consider now the probability density $P(\Delta h,
\ell)$ that the height difference between two points in the
reference plane separated by a length $\ell$ is within $d(\Delta h)$
of the height difference $\Delta h$. The anisotropic scaling
properties of the surface manifest themselves through the following
invariance under affine scale transformation
\begin{equation}
\lambda^H P(\lambda^H \Delta h, \lambda \ell) = P(\Delta h, \ell)
\label{def_affine} \ ,
\end{equation}
where $H$ is the roughness exponent. This is the starting point for
the discussion of self affine properties of fracture surfaces. Later
we will see that this definition is too limited and cannot capture
the rich complexity exhibited by fracture surface morphology. The
first part of this paper focuses on two dimensional fracture where
the generated rupture lines are 1+1 dimensional graphs. In Sec.
\ref{multiscaling} we show that the statistical properties of
rupture lines cannot be fully characterized by the scaling
invariance of Eq. (\ref{def_affine}). In that case, the more complex
structure of the probability distribution function leads to {\em
multiscaling} in contrast with {\em monoscaling} implied by Eq.
(\ref{def_affine}). We emphasize the properties of 1+1 dimensional
disordered fracture needed to be explained by a proper theory. Sec.
\ref{theoretic} offers a short critical review of existing
theoretical approaches to the problem. We discuss the limitations of
these approaches and explain why, in our opinion, they do not
provide a satisfactory description of the underlying physics. In
Sec. \ref{model1} we describe a new theoretical model for the growth
of a crack in a two dimensional medium in the presence of material
disorder. We elaborate on the mathematical foundations of the model
based on a recent development in which the method of iterated
conformal maps was applied to the problem of elasticity in the
presence of irregular crack geometries. We summarize the results of
the model and show that they meet the basic requirements of Sec.
\ref{multiscaling}.

The second part of the paper discusses three dimensional fracture
where the generated surfaces are 2+1 dimensional graphs. In Sec.
\ref{SO2} we show how the scaling properties of fracture surfaces
can be rationalized by decomposing the height-height structure
function into the irreducible representations of the $SO(2)$
symmetry group. This method offers a new way of understanding the
anisotropic properties of fracture surfaces in the plane of
fracture. We propose new experiments in which the rotational
symmetry is not broken such that the scaling properties should be
particularly simple. Sec. \ref{Summary} offers a summary and
outlines for future research directions.

\section{Multiscaling in 1+1 Dimensional Fracture}
\label{multiscaling}

In 1+1 dimensions one denotes the graph as $h(x)$, where $h$ is the
height of the surface at point $x$ relative to a smooth reference
line and considers the $n^{th}$ order structure function
$S_n(\ell)$,
\begin{equation}
S_n(\ell)\equiv \langle |h(x+\ell)-h(x)|^n\rangle \ , \label{Sn}
\end{equation}
where angular brackets denote an average over all $x$. These
quantities are invariant under {\em affine} scale transformations
if they are homogeneous functions of their arguments
\begin{equation}
S_n(\lambda \ell)\sim \lambda^{\zeta^{(n)}} S_n(\ell) \ .
\label{zetan}
\end{equation}
If $\zeta^{(n)}$ is linearly related to $n$ the scaling properties
of the graph are called ``normal" and the graph is statistically
self-affine. On the other hand, if $\zeta^{(n)}$ is a non-linear
function of $n$, the structure functions are multiscaling (or
anomalous) and the graph under consideration is called
``multiaffine''. To our best knowledge, all the studies regarding
fracture in 1+1 dimensions treated the rupture lines as normal
graphs, implying that $\zeta^{(n)}=nH$ where $H$ is the roughness
(Hurst) exponent. Note that $H>1/2$ indicates the existence of {\em
positive correlations} in the process generating the surface which
implies that an upward incremental deviation (relative to the smooth
reference line) of the rupture line is more likely to be followed by
an upward deviation than an downward one and vice versa
\cite{88Fed}. This feature shows that the higher $H$ is, the
smoother is the surface. Experimental analyses of rupture lines in
quasi two-dimensional materials yielded a roughness exponent $H$
whose numerical value was close to $0.67$ \cite{92PABT, 93KHW,
94EMHR, 03SAN}, suggesting some universality of the surface
generating process. The proximity of this numerical value to the
exact ratio $H=2/3$, characterizing the roughness of directed
polymers in random media \cite{95BS}, has lead some authors to
suggest that the two problems are in the same universality class
\cite{91HHRa, 93KHW, 95H-HZ, 95BS}. In this view, fracture is
considered as a global minimization problem.

In a recent work \cite{05BouP} we have shown that the phenomenology
is much richer and fracture lines are multiscaling. An example that
provides us with information of sufficient accuracy to establish the
multiscaling characteristics is rupture lines in paper. The data
acquired by Santucci et al. \cite{04SVC} was obtained in experiments
where centrally notched sheets of fax paper were fractured by
standard tensile testing machine. Four resulting crack profiles
$h(x)$ were digitized. Each digitization contained a few thousand
points, where care was taken to insure that the smallest separation
between points in $x$ is larger than the typical fiber width; this
is important to avoid the artificial introduction of overhangs that
destroy the graph property.

\begin{figure}[here]
\centering \epsfig{width=.45\textwidth,file=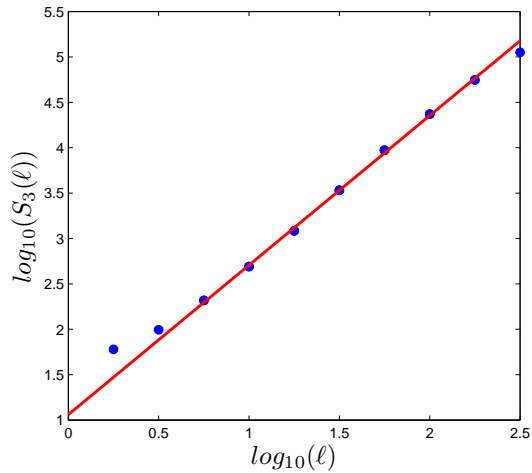} \caption{A
log-log plot of $S_3(\ell)$ as a function of $\ell$. The linear fit
corresponds to a typical scaling range of about $1.5$ orders of
magnitude, with a slope of $\zeta^{(3)} \simeq 1.65$.}
\label{Example}
\end{figure}

\begin{figure}[here]
\centering \epsfig{width=.45\textwidth,file=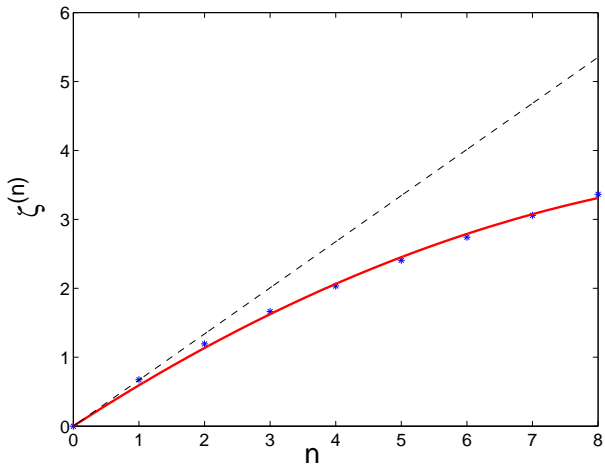}
\caption{The spectrum $\zeta^{(n)}$ as a function of the moment
order $n$ for rupture lines in paper. The function is fitted to the
form $\zeta^{(n)} = nH-n^2\lambda$ and the parameters $H$ and
$\lambda$ are given. The errors in the estimation of these
parameters reflect both the variance between different samples and
the fit quality. The linear plot $n\zeta^{(1)}$ is added to stress
the non-linear nature of $\zeta^{(n)}$.}\label{Multiaffine_spectrum}
\end{figure}

Denoting $\Delta h(\ell) \equiv h(x+\ell)-h(x)$ we analyzed the
data by boxing $\ell$ in logarithmic boxes, accumulating the data
between $10^0$ and $10^{0.25}$ (the smallest box) and between
$10^{2.25}$ and $10^{2.5}$ (the largest box). The $m^{th}$ box was
considered as representing data for $\ell=10^{m\times 0.25}$. On
the basis of this boxing we constructed the probability
distribution function (pdf) $P(\Delta h(\ell))$, which is just the
probability distribution function defined in Eq.
(\ref{def_affine}), by combining data from all the four samples.
Samples that exhibit marked trends (probably due to the finite
size of the sample), were detrended by subtracting the mean from
each distribution. The computed pdf's were then used to compute
the moments (\ref{Sn}), and these in turn, once presented as
log-log plots, yield the scaling exponents $\zeta^{(n)}$. Such a
typical log-log plot is shown in Fig. \ref{Example}, exhibiting a
typical scaling range of more than $1.5$ orders of magnitude. The
resulting values of the scaling exponents $\zeta^{(n)}$ are shown
in Fig. \ref{Multiaffine_spectrum}.

\begin{figure}[here]
\centering \epsfig{width=.475\textwidth,file=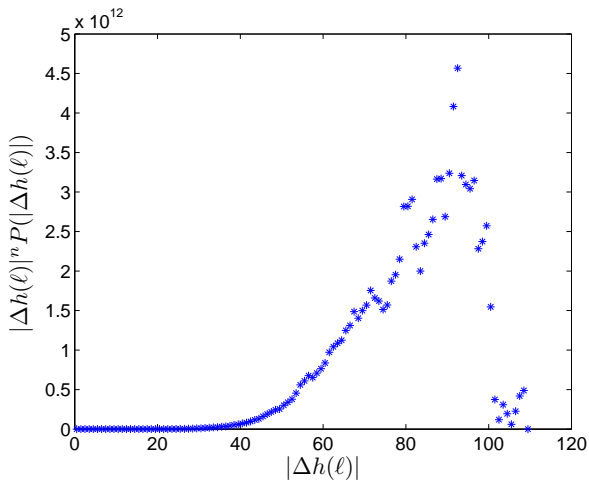}
\caption{An example of the convergence of the integral in Eq.
(\ref{integral}) for $\ell=10^{2.25}$ and $n=8$.}
\label{Convergence}
\end{figure}

As a function of $n$ these numbers  can be fitted to the quadratic
function $\zeta^{(n)} = nH-n^2\lambda$ with $H = 0.64 \pm 0.03$
and $\lambda = 0.026 \pm 0.004$ (a linear plot $n \zeta^{(1)}$ is
added for reference). The error bars quoted here reflect both the
variance between different samples and the fit quality.  The
exponents were computed for $n\le 8$ since for higher moments the
discrete version of the integral
\begin{equation}
\int |\Delta h(\ell) |^n P(\Delta h(\ell))d \Delta h(\ell)
\label{integral}
\end{equation}
did not converge. On the other hand, the convergence of the $8^{th}$
order moment is demonstrated in Fig.~\ref{Convergence}.

\begin{figure}
\centering \epsfig{width=.48\textwidth,file=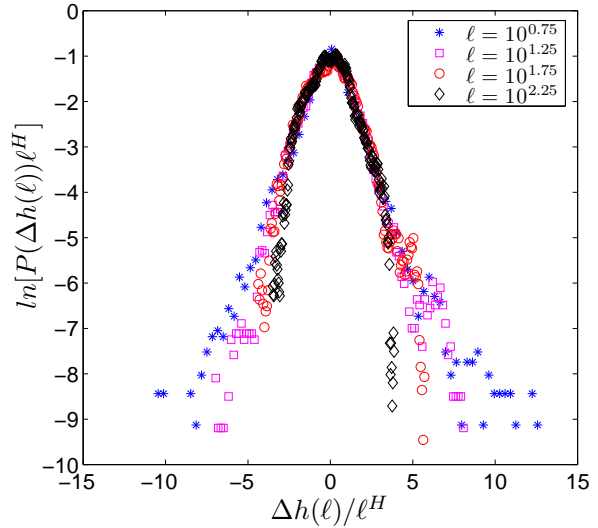}
\caption{The natural logarithm of $P(\Delta h(\ell))\ell^{H}$ as a
function of $\Delta h(\ell)/\ell^{H}$ for $H=0.64$. The legend gives
the scale $\ell$ for each plot.}\label{Data_Collapse}
\end{figure}

The point to stress is that the scaling exponents $\zeta^{(n)}$
depend non-linearly on $n$; for the range of $n$ values for which
the moments converge, the exponents can be fitted to a quadratic
function. It is well known from other areas of nonlinear physics,
and turbulence in particular \cite{84AGHA,95Fri,05BP}, that such
phenomena of multiscaling are associated with pdf's on different
scales $\ell$ that cannot be collapsed by simple rescaling. In
other words, in the absence of multiscaling, there exists a single
scaling exponent $H$ with the help of which one can rescale the
pdf's according to
\begin{equation}
P(\Delta h(\ell)) \sim \ell^{-H} f\left(\frac{\Delta
h(\ell)}{\ell^{H}}\right)\ , \label{distribution}
\end{equation}
where $f(\cdot)$ is a scaling function. In our case such rescaling
does not result in data collapse. In Fig. \ref{Data_Collapse} the
natural logarithm of $P(\Delta h(\ell))\ell^{H}$ is plotted as a
function of $\Delta h(\ell)/\ell^{H}$ for $H=0.64$. Indeed, the data
does not collapse onto a single curve. The fatter tails of the
probability distribution functions at smaller scales are typical to
multiscaling situations, where the statistics of rare events plays
an important role.

The multiscaling property of 1+1 dimensional fracture surfaces has
an important implication; Since the problem of directed polymers in
random media is {\em monoscaling} \cite{91Hal}, we conclude that the
two problems are not in the same universality class. In fact none of
the known kinetic growth model of statistical physics \cite{95BS} is
consistent with the multiscaling spectrum discovered in 1+1
dimensional fracture. This might suggest that 1+1 dimensional
fracture defines a distinct universality class in statistical
physics, a proposition that must be checked against a much broader
experimental data set. To conclude, the presence of multiscaling
offers a stringent test for any theoretical model of 1+1 dimensional
in-plane fracture.

\section{Traditional Theoretical Approaches}
\label{theoretic}

In the last fifteen years or so several statistical physics
approaches were applied in this field. In this section we offer a
short critical review of these approaches.

\subsection{Lattice Models}

The most studied model aiming at understanding the origin of
self-affinity of fracture surfaces morphology, the value of the
roughness exponent $H$ and its degree of universality is the
quasi-static random fuse model \cite{book}. This model consists of
an array of identically conducting linear (Ohmic) resistors placed
between two electrical bus bars with a fixed voltage difference. A
threshold value $t_j$ taken from an uncorrelated distribution $p(t)$
is assigned to each resistor. At each step of the numerical simulation
the Kirchhoff equations are solved and the current $i_j$ passing
through each resistor is calculated. The system is advanced in time
according to the extremal rule that states that the resistor with
the largest ratio between current and threshold $max_j(i_j/t_j)$ is
irreversibly removed. The system is driven to its final state where
the network of fuses stops conducting. The behavior of this model is
controlled by the width of the threshold distribution $p(t)$
\cite{91HHRb}. For ``very strong'' disorder, i.e. for broad enough
threshold distributions, the model behaves like a percolation problem,  without
any  localized failure mode. For this regime, the
disorder dominates the correlations which are induced by current enhancements
due to removed resistors (the analog of stress concentration in the
elastic problem).  The cluster of broken bonds (removed resistors)
shows no anisotropic scaling (formally, the roughness exponent $H$
is one in this case). For smaller disorders damage is distributed
throughout the system, but at the ``critical phase'' of the dynamics
of the model a localized failure mode sets in, resulting in a
self-affine rupture line. At even smaller disorders, in the ``weak''
disorder regime, the system is ``critical'' from the early stages of
the evolution in the sense that a large crack quickly nucleates and
the dynamics is controlled by its propagation. The roughness exponent
$H$ for the regimes where self-affine rupture lines are generated
was measured to be $H \approx 0.7$ \cite{91HHRa, 00SRA}.

Although the random fuse model is widely regarded ``a paradigm for
brittle fracture'' \cite{03HS}, its capability to describe a real
physical system that fails under external load is questionable; All
the data on fracture surfaces morphology were obtained by in-plane
fracture experiments, mainly in mode I, whereas the random fuse
model is a discrete version of out-of-plane fracture (mode III),
which is unstable experimentally \cite{unstable}. While in-plane
fracture modes entail the solution of the fully {\em tensorial}
Lam\'e elasticity theory, out-of-plane fracture entails the solution
of {\em scalar} elasticity theory, described by the Laplace
equation. There is no a-priori reason to believe that the two
problems are in the same universality class. Moreover, the random
fuse model misrepresents another important feature of fracture
experiments; when a material is externally loaded intrinsic defects
can lead to the nucleation of a crack everywhere in the system.
Since this process occurs in an uncontrolled way, it is an
experimental convention to introduce a small crack in the sample
prior to loading. Typically, the stress concentration near the tip
of such a crack is so overwhelming that damage far from this region
cannot develop significantly even in heterogenous materials. This is
in contrast to the random fuse model, excluding situations of very
narrow disorders, where damage is accumulated throughout the system
{\em prior} to the critical phase of macroscopic failure in which
the final rupture line develops. In spite of the inadequacy of the
random fuse model to faithfully describe real physical systems, it
might be useful to understand how positive correlations are built in
the model, leading to $H>1/2$. Recently \cite{03HS} it was proposed
that a correlated gradient percolation process is responsible for
the value of the roughness exponent measured in this model. Here we
do not give the argument in full detail, but focus on the main
conceptual aspect of this proposition: correlations in the model are
crucially dependent on the damage accumulated in the system before
the critical phase of the formation of the final rupture line. In
this view, the positive correlations observed in the final rupture
line is a manifestation, via a coalescence process, of the long
range positive correlations that are built in the system during the
earlier stages of the dynamics. Therefore, this explanation rely
heavily on the presence of a broad enough threshold distribution
that is responsible for correlations in the precursory phase of the
dynamics. On the other hand, other works \cite{04NSZ, 05RA} support
the view that correlations in the damage accumulated prior to the
critical phase of macroscopic failure are {\em negligible}. If this
view is valid then the origin of self-affine crack roughness in the
random fuse model does not depend on whether there is ``strong'' or
``weak'' disorder since correlations are built in the system only at
the final stage of macroscopic failure. We think that this view
represents better the typical experimental situation where stress
concentration near the pre-existing crack dominates the failure
process. This view will be further developed below, see Sec.
\ref{model1}.

A second group of lattice models originated from the central force
model \cite{book}. In this model the lattice is connected by elastic
springs that are freely rotating around the nodes. This model is
much closer to real fracture since its continuum limit is the
tensorial Lam\'e elasticity. Unfortunately, no systematic studies of
the roughness exponent for this model were reported. A third group
of lattice models is the random beam models. In these models the
lattice is composed of elastic beams that are rigidly connected to
the nodes. Here the bond bending elasticity is taken into account by
considering the local rotational degrees of freedom at each node in
addition to the translational ones that are described by the central
force model. The roughness exponent in 1+1 dimensions for this model
was found to be $H \approx 0.86$ \cite{01SHH}. Therefore, this model
seems to be unrelated to the experimental findings. The origin of
this discrepancy is probably the fact that Lam\'e elasticity
provides an appropriate description of materials on a large enough
length scale. To conclude, we propose that none of the lattice
models found in the literature are capable of describing 1+1
dimensional in-plane fracture and to faithfully represent the
experiments in the field.

\subsection{Continuum Models}

The basic issue arising in continuum models of crack propagation is
how to predict where the crack goes under the action of a given
stress field. One can use symmetry principles to derive an equation
of motion for the crack tip. Adapting Eq. (\ref{sigma}) to
two-dimensional quasi-static fracture, the stress tensor field
attains the universal form
\begin{equation}
\sigma_{ij}(r,\theta)=\frac{K_{_{\rm I}}}{\sqrt{2\pi
r}}\Sigma^I_{ij}(\theta)+ \frac{K_{_{\rm II}}}{\sqrt{2\pi
r}}\Sigma^{II}_{ij}(\theta) \  , \label{universalform}
\end{equation}
where $I$ and $II$ denote the mode I (tensile) and mode II (shear)
parts of the stress tensor field. As was explained above, the $r^{-1/2}$ singularity is not physical;
nevertheless in many cases there exists a cut-off distance from the crack tip (a
one-dimensional front), above which the
stress field is dominated by the universal form. Therefore, the
non-universal parts, the so-called stress intensity factors
$K_{_{\rm I}}$ and $K_{_{\rm II}}$ are very important physical
quantities.

In order to derive an equation of motion for the crack
tip  consider a curved crack in a two
dimensional medium. The position of the crack tip is ${\bf r}^{tip}$
in some convenient coordinate system. Let us denote the unit tangent and the
unit normal to the curved crack at its tip by $\hat{\bf t}$ and
$\hat{\bf n}$ respectively and assume that the crack tip fields
are characterized by the stress intensity factors $K_{_{\rm I}}$
and $K_{_{\rm II}}$. A discrete symmetry that the crack tip
dynamics should remain invariant under its operation is $\hat{\bf n} \rightarrow -\hat{\bf n}$. Under this
symmetry the relevant quantities in the problem transform as
follows: (i) $K_{_{\rm I}} \rightarrow K_{_{\rm I}}$ (ii)
$K_{_{\rm II}} \rightarrow -K_{_{\rm II}}$ (iii) $v \rightarrow
v$. Now one can write down the most general first order
differential equations that are invariant under this symmetry. The first
equation is just the kinematic relation for the rate of crack growth
\begin{equation}
\frac{\partial {\bf r}^{tip}}{\partial t} =  v(K_{_{\rm
I}},K^2_{_{\rm II}})~\!\hat{\bf t} \ .
\end{equation}
The second one describes the rate of crack tip rotation
\begin{equation}
\label{HS}
 \frac{\partial \hat{\bf t}}{\partial t}  =  -
f(K_{_{\rm I}},K^2_{_{\rm II}}, v) K_{_{\rm II}} \hat{\bf n} \ ,
\end{equation}
where $f(K_{_{\rm I}},K^2_{_{\rm II}}, v)$ is a positive material
function. These equations were originally derived in \cite{93HS}
and were shown to be consistent with various experimental
observations in \cite{03BHP, 05BMP}. To further simplify Eq.
({\ref{HS}) we rewrite the local crack tip directions in terms of
the angle $\theta$ that the unit tangent makes with the x-axis as
\begin{eqnarray}
\hat \B t&=&\cos{\theta}\hspace{.1cm}\bf\hat{x}+\sin{\theta}\hspace{.1cm}\bf\hat{y}\nonumber\\
\hat  \B
n&=&-\sin{\theta}\hspace{.1cm}\bf\hat{x}+\cos{\theta}\hspace{.1cm}\bf\hat{y}
\ .
\end{eqnarray}
Using these expressions Eq. (\ref{HS}) can be written as
\begin{equation}
\label{rotation} \frac{\partial\theta}{\partial t}=-f(K_{_{\rm
I}},K^2_{_{\rm II}})K_{_{\rm II}} \ .
\end{equation}

In order to use this equation, consider a long crack propagating
quasi-statically in a two dimensional medium under mode I
conditions. If the medium were homogenous, we would have $K_{_{\rm
II}}\{h(x)\} = 0$, and according to Eq. ({\ref{rotation}) the crack
would extend along a straight path. Real material possess however
some inhomogeneities which induces fluctuations $h(x)$ in the crack
path. The global symmetry is broken locally, leading to $K_{_{\rm
II}}\{h(x)\} \ne 0$. $K_{_{\rm II}}\{h(x)\}$ is a functional of the
fluctuations $h(x)$, typically including a long-range contribution,
reflecting the long-range nature of the elastic interactions and a
local contribution. If we restrict ourselves to small fluctuations
$h(x)$ we obtain $\theta \approx \partial_{x}h$. Furthermore, by
considering steady state propagation such that $\partial_{t}
\rightarrow  v\partial_{x}$ and representing the effect of material
disorder by a noise term $\vartheta(x)$, we arrive at the following
Langevin type equation
\begin{equation}
\frac{v}{f}~\partial_{xx}h(x) = -K^{(1)}_{_{\rm II}}\{h(x)\} -
\vartheta(x) \ , \label{Langevin}
\end{equation}
where the superscript $(1)$ denotes the fact the $K_{_{\rm
II}}\{h(x)\}$ is considered to first order in the small
fluctuation $h(x)$ and the disorder is modeled as
uncorrelated noise
\begin{equation}
\langle \vartheta(x) \vartheta(x') \rangle \sim \delta(x-x') \ .
\label{noise}
\end{equation}
The mode II stress intensity factor, $K_{_{\rm II}}\{h(x)\}$, was
calculated in first order perturbation theory in \cite{80CR},
yielding
\begin{equation}
K^{(1)}_{_{\rm II}}\{h(x)\} = \!\frac{1}{2} K^{(0)}_{_{\rm I}}
\partial_x h + \sqrt{\frac{2}{\pi}}\int_{-\infty}^{x}\!\!
\frac{\partial_{x'}(h(x')\sigma^{(0)}_{xx}(x'))}{\sqrt{-x'}}dx'
\!\!\ , \label{perturb}
\end{equation}
where the superscript $(0)$ denotes quantity of the unperturbed
solution. By considering a configuration in which
$\sigma^{(0)}_{xx}(x')$ is practically constant, a scaling analysis
of Eq. (\ref{Langevin}) results in a logarithmic roughness of the
rupture line. A similar treatment as the one presented here was
developed in \cite{97REF} where higher order terms were shown to be
irrelevant in the renormalization group sense. It is concluded that
this approach does not produce any non-trivial roughness exponent.
The reason for this failure is, in our opinion, the fact that the
crack growth takes place {\em at the tip} which is controlled
completely by the stress intensity factor $K_{_{\rm II}}$. We will
show that there are observations that motivate the view that the
crack growth is a discrete process taking place via the nucleation
of damage at {\em a finite distance $R$ ahead of the crack tip}.
This length $R$ characterizes the scale of ``spikiness'' of the
crack path; thus, when the growth step takes place at a distance $R$
ahead of the crack tip the irregularities of the crack path
(``spikiness'') change the structure of the stress field and a
different long range quantity controls the growth process. Moreover,
the irregularities of the crack path are difficult to handle using
perturbative approaches. The effect of this finite length scale on
the crack growth is further discussed in the next section. We
conclude this section by noting that the origin of the long range
positive correlations in 1+1 dimensional fracture and its
multiscaling nature are not yet explained by the existing models.

\section{A Model}
\label{model1}

In this section we review a recently introduced statistical model of
crack surfaces morphology in 1+1 dimensional fracture which takes
into account the insights gained from both experiments and previous
theoretical approaches. It is common to divide the failure of
materials into two categories: ductile and brittle. In brittle
fracture materials break when subjected to large enough stresses
without large plastic deformations taking place. In ductile
fracture, on the other hand, materials undergo significant plastic
flow before experiencing any local failure. In this case a crack
evolves by the nucleation, growth and coalescence of voids. A recent
experiment  \cite{03C} motivated the view that quasi-static brittle
fracture occurs, similarly to ductile fracture, by the nucleation of
damage ahead of the crack tip. It is not clear yet if this process
necessarily involves plastic deformations. The length scale of
damage nucleation $R$ in traditional ``brittle'' and ``ductile''
materials is completely different. Nevertheless, we suggest that
this behavior is {\em generic} irrespective of whether the physics
behind this length scale is related to plastic deformation or to
existing material disorder.

Therefore, a statistical theory of crack surfaces morphology in 1+1
dimensions should include the following components: (i) the exact
solution of the elasticity problem in the presence of irregular
crack geometries (ii) a growth law in which the evolution of the
crack is controlled by the nucleation of voids at a {\em finite
distance} $R$ ahead of the crack tip (iii) an appropriate
description of the effect of material disorder.

\subsection{The elasticity problem in the presence of irregular crack
geometries}

In this section we review a non-perturbative approach to the
calculation of the stress field in two space dimensions for an
arbitrarily shaped crack, based on conformal mapping \cite{04BMPa}.
We start with the quasi-static version of Eq. (\ref{EOM}), obtained
by dropping the inertial term on the right hand side
\begin{equation}
\frac{\partial \sigma_{ij}}{\partial x_j}= 0 . \label{equil}
\end{equation}
For in-plane modes of fractures in two dimensions one introduces the
Airy stress potential $U(x,y)$ such that
\begin{equation}
\sigma_{xx}=\frac{\partial^2 U}{\partial y^2} \ ; \sigma_{xy}=-
\frac{\partial^2 U}{\partial x\partial y} \ ;
\sigma_{yy}=\frac{\partial^2 U}{\partial x^2} \ . \label{sigU}
\end{equation}
Thus the set of Eq. (\ref{equil}), after simple manipulations,
translate to a bi-Laplace equation for the Airy stress potential
$U(x,y)$ \cite{86LL}
\begin{equation}
\nabla^2 \nabla^2 U(x,y) =0 \ , \label{bilaplace}
\end{equation}
with the prescribed boundary conditions on the crack and on the
external boundaries of the material. At this point we choose to
focus on the case of uniform remote loadings and traction-free crack
boundaries. This choice, although not the most general, is of great
interest and will serve to elucidate our method. Other solutions may
be obtained by superposition. Thus, the boundary conditions at
infinity, for the two in-plane symmetry modes of fracture, are
presented as
\begin{eqnarray}
&&\!\!\!\!\!\!\!\!\sigma_{xx}(\infty)=0\ ;
\sigma_{yy}(\infty)=\sigma_\infty\ ;
 \sigma_{xy}(\infty)=0\quad \text{Mode I}\label{mode1}\\
&&\!\!\!\!\!\!\!\!\sigma_{xx}(\infty)=0\ ; \sigma_{yy}(\infty)=0\ ;
 \sigma_{xy}(\infty)=\sigma_\infty\quad \text{Mode II}\ . \nonumber\\ \label{mode2}
\end{eqnarray}
In addition, the free boundary conditions on the crack are expressed
as
\begin{equation}
\sigma_{xn}(s)=\sigma_{yn}(s)=0  \ , \label{bcm12}
\end{equation}
where $s$ is the arc-length parametrization of the crack boundary
and the subscript $n$ denotes the outward normal direction
$\hat{{\bf n}}$.

The solution of the bi-Laplace equation can be written in terms of
{\em two} analytic functions $\phi(z)$ and $\eta(z)$ as
\begin{equation}
U(x,y)= \Re [\bar z\varphi(z)+\eta(z)] \ . \label{Uphichi}
\end{equation}
In terms of these two analytic functions, using Eq. (\ref{sigU}),
the stress components are given by
\begin{eqnarray}
\sigma_{yy}(x,y)&=&\Re [2 \varphi'(z)+ \bar z\varphi''(z)+\eta''(z)]\nonumber\\
\sigma_{xx}(x,y)&=&\Re [2 \varphi'(z)-\bar z\varphi''(z)-\eta''(z)]\nonumber\\
\sigma_{xy}(x,y)&=&\Im [\bar z\varphi''(z)+\eta''(z)].
\label{components}
\end{eqnarray}
In order to compute the full stress field one should first formulate
the boundary conditions in terms of the analytic functions
$\varphi(z)$ and $\eta(z)$ and to remove the gauge freedom in Eq.
(\ref{Uphichi}). The boundary conditions Eq. (\ref{bcm12}), using
Eq. (\ref{sigU}), can be rewritten as \cite{53Mus}
\begin{equation}
\partial_s\left[\frac{\partial U}{\partial x}
+i\frac{\partial U}{\partial y}\right]=0\ . \label{bcU}
\end{equation}
Note that we do not have enough boundary conditions to determine
$U(x,y)$ uniquely. In fact we can allow in Eq. (\ref{Uphichi})
arbitrary transformations of the form
\begin{eqnarray}
\varphi &\rightarrow& \varphi +iCz+\gamma\nonumber\\
\psi &\rightarrow& \psi +\tilde\gamma \ , \quad\psi\equiv \eta'
\end{eqnarray}
where $C$ is a real constant and $\gamma$ and $\tilde\gamma$ are
complex constants. This provides five degrees of freedom in the
definition of the Airy potential. Two of these freedoms are removed
by choosing the gauge in Eq. (\ref{bcU}) according to
\begin{equation}
\frac{\partial U}{\partial x} +i\frac{\partial U}{\partial y} = 0  \
, \quad \text{on the boundary}\ . \label{choice}
\end{equation}
It is important to stress that whatever the choice of the five
freedoms, the stress tensor is unaffected; see \cite{53Mus} for an
exhaustive discussion of this point. Computing Eq. (\ref{choice}) in
terms of Eq. (\ref{Uphichi}) we arrive at the boundary condition
\begin{equation}
\varphi(z)+z\overline{\varphi'(z)}+\overline{\psi(z)}=0 \ .
\label{bccrack}
\end{equation}
To proceed we represent $\varphi(z)$ and $\psi(z)$ in Laurent
expansion form:
\begin{eqnarray}
\varphi(z) &=&\varphi_1 z + \varphi_0
+\varphi_{-1}/z+\varphi_{-2}/z^2+\cdots \ , \nonumber\\
\psi(z) &=&\psi_1 z + \psi_0 +\psi_{-1}/z+\psi_{-2}/z^2+\cdots \ .
\label{Laurentpp}
\end{eqnarray}
This form is in agreement with the boundary conditions at infinity
that disallow higher order terms in $z$. One freedom is now used to
choose $\varphi_1$ to be real and two more freedoms will allow us
later on to fix $\varphi_0$. Then, using the boundary conditions
(\ref{mode1}) and (\ref{mode2}), we find
\begin{eqnarray}
 \varphi_1&=&\frac{\sigma_{\infty}}{4}\ ;\quad
 \psi_1=\frac{\sigma_{\infty}}{2} \quad  \text{ Mode I}  \ ,\nonumber\\
 \varphi_1&=&0 \ ; \quad\quad \psi_1=i\sigma_\infty
\quad \text{ Mode II} \ . \label{p1p1}
\end{eqnarray}

%

As said above, the direct determination of the stress tensor for a
given arbitrary shaped crack is difficult. To overcome the
difficulty we perform an intermediate step of determining the
conformal map from the exterior of the unit circle to the exterior
of our given crack. Before our work the best available approach for such a
task was the Schwartz-Cristoffel transformation. We have  presented
an alternative new approach for finding the wanted conformal
transformation, given in terms of a functional iteration of
fundamental conformal maps. The use of iterated conformal maps was
pioneered by Hastings and Levitov \cite{98HL}; it was subsequently
turned into a powerful tool for the study of fractal and fracture
growth patterns \cite{99DHOPSS,00DFHP,00DLP,01BJLMP,01BDLP,02BLP,05MPST}.
In the next subsection we describe how, given a crack shape, to
construct a conformal map from the complex $\omega$-plane to the
physical $z$-plane  such that the conformal map $z = \Phi(\omega )$
maps the exterior of  the unit circle in the $\omega$-plane to the
exterior of the crack in the physical $z$-plane, after $n$ directed
growth steps. We draw the reader's attention to the fact that this
method is more general than its application in this paper \cite{05MPST}, and in
fact we offer it as a superior method to the Schwartz-Cristoffel
transformation, with hitherto undetermined potential applications in
a variety of two-dimensional contexts.

\begin{figure}
\centering \epsfig{width=.45\textwidth,file=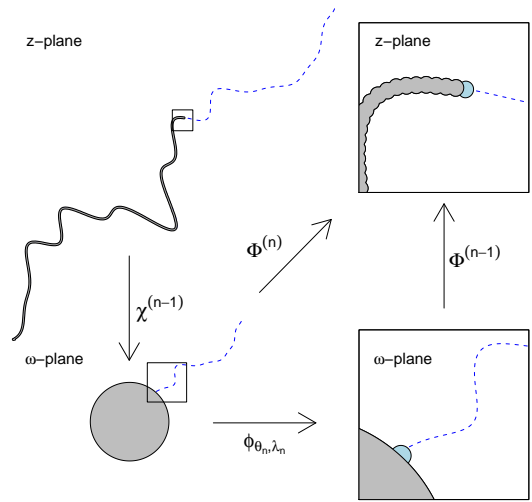}
\caption{Example of how to construct the conformal mapping along a
line, see text for details.}\label{method}
\end{figure}

\subsubsection{The conformal mapping}

The essential building block in the present application, as in all
the applications of the method of iterated conformal maps is the
fundamental map $\phi_{\lambda,\theta}$ that maps the exterior
circle onto the unit circle with a semi-circular bump of linear size
$\sqrt\lambda$ which is centered at the point $e^{i\theta}$. This
map reads \cite{98HL}:
\begin{eqnarray}
\label{phi}
   &&\phi_{0,\lambda}(w) = \sqrt w \left\{ \frac{(1+
   \lambda)}{2w}(1+w)\right. \\
   &&\left.\times \left [ 1+w+w \left( 1+\frac{1}{w^2} -\frac{2}{w}
\frac{1-\lambda} {1+ \lambda} \right) ^{1/2} \right] -1 \right \} ^{1/2} \nonumber\\
   &&\phi_{\theta,\lambda} (w) = e^{i \theta} \phi_{0,\lambda}(e^{-i
   \theta}
   w) \,.
   \label{eq-f}
\end{eqnarray}
The inverse mapping $\phi^{-1}_{\theta=0,\lambda}$ is of the form
\begin{equation}
  \label{eq:3}
  \phi^{-1}_{0,\lambda}= \frac{\lambda z-\sqrt{1+\lambda}(z^2-1)}{1-(1+\lambda)z^2}z \ .
\end{equation}
By composing this map with itself $n$ times with a judicious choice
of series $\{\theta_k\}_{k=1}^n$ and $\{\lambda_k\}_{n=1}^{n}$ we
will construct $\Phi^{(n)}(\omega)$ that will map the exterior of
the circle to the exterior of an arbitrary simply connected shape.
To understand how to choose the two series $\{\theta_k\}_{k=1}^n$
and $\{\lambda_k\}_{n=1}^{n}$ consider Fig. \ref{method} and define
the inverse map $\omega=\chi^{(n)}(z)$.  Assume now that we already
have $\Phi^{(n-1)}(\omega)$ and therefore also its analytic inverse
$\chi^{(n-1)}(z)$ after $n-1$ growth steps and we want to perform
the next iteration. To construct $\Phi^{(n)}(\omega)$ we advance our
mapping in the direction of a point $\tilde z$ in the $z$-plane by
adding a bump in the direction of $\tilde w=\chi^{(n-1)}(\tilde z)$
in the $w$-plane. The map $\Phi^{(n)}(\omega)$ is obtained as
follows:
\begin{equation}
\label{conformal} \Phi^{(n)}(\omega) =
\Phi^{(n-1)}(\phi_{\theta_n,\lambda_n}(\omega )) \ . \label{iter}
\end{equation}
The value of $\theta_n$ is determined by
\begin{equation}
  \label{eq:1}
  \theta_n=\arg [\chi^{(n-1)}(\tilde z)]
\end{equation}
The magnitude of the bump $\lambda_n$ is determined by requiring
fixed size bumps in the $z$-plane. This means that
\begin{equation}
  \label{eq:2}
  \lambda_n = \frac{\lambda_0}{|{\Phi^{(n-1)}}' (e^{i \theta_{n}})|^2}.
\end{equation}
We note here that it is not necessary in principle to have fixed
size bumps in the physical domain. In fact, adaptive size bumps
could lead to improvements in the precision and performance of our
scheme. We consider here the fixed size scheme for the sake of
simplicity and we will show that the accuracy obtained is sufficient
for our purposes. Iterating the scheme described above we end up
with a conformal map that is written in terms of an iteration over
the fundamental maps (\ref{phi}):
\begin{equation}
  \label{eq:3}
  \Phi^{(n)}(w)=\phi_{\theta_{1},\lambda_{1}}\circ\ldots\circ\phi_{\theta_n,\lambda_n}(w) \ .
\end{equation}
For the sake of newcomers to the art of iterated conformal maps we
stress that this iterative structure is abnormal, in the sense that
the order of iterates in inverted with respect to standard dynamical
systems. On the other hand the inverse mapping follows a standard
iterative scheme
\begin{equation}
  \label{eq:4}
\chi^{(n)}(z)=\phi^{-1}_{\theta_{n},\lambda_{n}}\circ\ldots\circ\phi^{-1}_{\theta_1,\lambda_1}(z)
\ .
\end{equation}

The algorithm is then described as follows; first we divide the
curve into segments separated by points $\{z_i\}$. The spatial
extent of each segment is taken to be approximately
$\sqrt{\lambda_0}$, in order to match the size of the bumps in the
$z$-plane. Without loss of generality we can take one of these
points to be at the center of coordinates and to be our starting
point. From the starting point we now advance along the shape by
mapping the next point $z_i$ on the curve according to the scheme
described above.


\subsubsection{Solution in terms of conformal mappings}

The conformal map $\Phi^{(n)}(\omega)$ is constructed in $n$
iterative steps. For the discussion below we do not need the $n$
superscript and will denote simply $\Phi(\omega ) \equiv
{\Phi}^{(n)}(\omega )$. This map is univalent \cite{99DHOPSS},
having the Laurent expansion form
\begin{equation}
\Phi(\omega ) = F_1\omega + F_0
+F_{-1}/\omega+F_{-2}/\omega^2+\cdots \ . \label{Laurent}
\end{equation}
Any position $z$ in the physical domain is mapped by  $\chi(z)
\equiv {\Phi}^{-1}(z)$ onto a position $\omega$ in the mathematical
domain. This transformation does not immediately provide the
solution as the bi-Laplace equation, in contrast to the Laplace
equation, is not conformally invariant. Nevertheless, the conformal
mapping method can be extended to non-Laplacian problems. We begin
by writing our unknown functions $\varphi(z)$ and $\psi(z)$ in terms
of the conformal map
\begin{equation}
\varphi(z)\equiv \tilde \varphi\left(\chi(z)\right) \ , \quad
\psi(z)\equiv \tilde \psi\left(\chi(z)\right) \ . \label{ppp}
\end{equation}
Using the Laurent form of the conformal map, Eq. (\ref{Laurent}),
the linear term as $\omega\to \infty$ is determined by Eqs.
(\ref{ppp}). We therefore write
\begin{eqnarray}
\tilde \varphi(\omega) &=& \varphi_1F_1 \omega
+ \tilde\varphi_{-1}/\omega+\tilde\varphi_{-2}/\omega^2 +\dots\ , \nonumber\\
\tilde \psi(\omega) &=& \psi_1F_1 \omega + \tilde \psi_0
+\tilde\psi_{-1}/\omega +\tilde\psi_{-2}/\omega^2 +\dots\ ,
\label{expom}
\end{eqnarray}
where we used the last two freedoms to choose $\varphi_0=
-F_0\varphi_0$ such that $\tilde\varphi_0=0$. The boundary condition
(\ref{bccrack}) is now read for the unit circle in the
$\omega$-plane. Denoting $\epsilon\equiv e^{i\theta}$ and
\begin{equation}
u(\epsilon)\equiv \sum_{k=1}^\infty \tilde\varphi_{-k}/\epsilon^k \
, \quad v(\epsilon)\equiv  \sum_{k=0}^\infty
\tilde\psi_{-k}/\epsilon^k \ , \label{uvdef}
\end{equation}
we write
\begin{equation}
u(\epsilon) + \frac{\Phi(\epsilon)}{\overline{\Phi^{'}(\epsilon)}}
\overline{u'(\epsilon)}+\overline{v(\epsilon)}=f(\epsilon) \ .
\label{eq.fund}
\end{equation}
The function $f(\epsilon)$ is a known function that contains all the
coefficients that were determined so far:
\begin{equation}
f(\epsilon) =-\varphi_1 F_1\epsilon - \frac{\Phi(\epsilon)}
{\overline{\Phi^{'}(\epsilon)}}\varphi_1 F_1
-\frac{\overline{\psi_1} F_1}{\epsilon} \label{ff} \ .
\end{equation}
To solve the problem we need to compute the coefficients
$\tilde\varphi_n$ and $\tilde\psi_n$. To this aim we first write
\cite{02BLP}
\begin{equation}
\frac{\Phi(\epsilon)}{\overline{\Phi^{'}(\epsilon)}}=
\sum_{-\infty}^{\infty}b_i\epsilon^i. \label{expmap}
\end{equation}
The function $f(\epsilon)$ has also an expansion of the form
\begin{equation}
f(\epsilon)= \sum_{-\infty}^{\infty}f_{i}\epsilon^i \ .
\label{fepsilon}
\end{equation}
In the discussion below we assume that the coefficients $b_i$ and
$f_i$ are known. In order to compute these coefficients we need to
Fourier transform the function $\Phi(\epsilon)/
\overline{\Phi^{'}(\epsilon)}$. This is the most expensive step in
our solution. One needs to carefully evaluate the Fourier integrals
between the branch cuts. Using the last two equations together with
Eqs. (\ref{uvdef}) and (\ref{eq.fund}) we obtain
\begin{eqnarray}
\tilde \varphi_{-m}&-&\sum_{k=1}^\infty k~b_{-m-k-1}
\tilde\varphi^*_{-k} =f_{-m}
\ , \quad m=1,2\cdots \ , \label{power_sol1}\\
\tilde \psi^*_{-m}&-&\sum_{k=1}^\infty k~b_{m-k-1}
\tilde\varphi^*_{-k} =f_{m} \ . \quad m=0,1,2\cdots \
\label{power_sol2}
\end{eqnarray}
These sets of linear equations are well posed.  The coefficients
$\tilde \varphi_{-m}$ can be calculated from equation
(\ref{power_sol1}) alone,  and then they can be used to determine
the coefficients $\tilde \psi_{-m}$. This is in fact a proof that
Eq. (\ref{eq.fund}) determines the functions $u(\epsilon)$ and
$v(\epsilon)$ together. This fact had been proven with some
generality in \cite{53Mus}.

The calculation of the Laurent expansion form of $\tilde
\varphi(\omega)$ and $\tilde \psi(\omega)$ provides the solution of
the problem in the $\omega$-plane. Still, one should express the
derivatives of $\varphi(z)$ and $\eta(z)$ in terms of $\tilde
\varphi(\omega)$ and $\tilde \psi(\omega)$ and the inverse map
$\chi(z)$ to obtain the solution in the physical $z$-plane. This is
straight forward and yields
\begin{eqnarray}
\varphi'(z)&=&\tilde\varphi'[\chi(z)] ~\chi'(z)\nonumber\\
\varphi''(z)&=&\tilde\varphi''[\chi(z)] ~[\chi'(z)]^2+
 \tilde\varphi'[\chi(z)] ~\chi''(z)\nonumber\\
\eta''(z)&=& \psi'(z)=\tilde\psi'[\chi(z)] ~\chi'(z).
\label{relations}
\end{eqnarray}
Upon substituting these relations into Eq. (\ref{components}) one
can calculate the {\em full} stress field for an arbitrarily shaped
crack. The expression of the stress field in terms of the inverse
conformal mapping is known for quite a long time although it is very
limited as the conformal mapping and its inverse is rarely at hand.
The central step of progress here is the conjunction of the novel
functional iterative scheme for obtaining the inverse conformal
mapping with the known result that expresses the stress field in
terms of this inverse mapping. This method was shown in
\cite{04BMPa} to reproduce results for known regular crack
geometries.

\subsection{A Crack Growth Law}

After developing the mathematics needed for the calculation of the
linear elastic stress field for irregular crack geometries, we turn
now to a description of the crack growth law and the effect of
disorder. As was explained above, a crucial aspect of the crack
growth process is the finite length scale $R$ in which damage
nucleates. A simple model \cite{04BMPb, 05ABKMP} for $R$ can be
developed by assuming the near crack tip zone to be properly
described by the Huber-von Mises plasticity theory \cite{90Lub}.
This theory focuses on the deviatoric stress $s_{ij}\equiv
\sigma_{ij} -\case{1}{3}{\rm Tr} (\B\sigma)\delta_{ij}$ and on its
invariants. The second invariant, $J_2\equiv \case{1}{2}
s_{ij}s_{ij}$, corresponds to the distortional energy. The material
yields as the distortional energy exceeds a material-dependent
threshold $\sigma_{_{\rm Y}}^2$. The fact that we treat this
threshold as a constant, independent of the state of deformation and
its history, implies that we specialize for ``perfect'' plasticity.
In two-dimensions this yield condition reads \cite{90Lub}
\begin{equation}
J_2
=\frac{\sigma^2_{1}-\sigma_{1}\sigma_{2}+\sigma^2_{2}}{3}=\sigma^2_{_{\rm
Y}} \ . \label{Mises}
\end{equation}
Here $\sigma_{1\!,2}$ are the principal stresses given by
\begin{equation}
\sigma_{1\!,2} = \frac{\sigma_{yy}+\sigma_{xx}}{2} \pm
\sqrt{\frac{(\sigma_{yy}-\sigma_{xx})^2}{4}+\sigma_{xy}^2} \ .
\label{PrincipalStress}
\end{equation}

In the purely linear-elastic solution the crack tip region is where
high stresses are concentrated. Perfect plasticity implies on the
one hand that the tip is blunted and on the other hand that inside
the plastic zone the Huber-von Mises criterion (\ref{Mises}) is
satisfied. The outer boundary of the plastic zone will be called
below the ``yield curve'', and in polar coordinates around the crack
tip will be denoted ${\cal R}(\theta)$. Below we will compute the
outer stress field {\em exactly} as was explained in the previous
section. Using this field we can find the yield curve ${\cal
R}(\theta)$. Such a typical curves are shown in the insets of Figs.
\ref{straightsteps} and \ref{pdf}.

Whatever is the actual shape of the blunted tip its boundary cannot
support stress. Together with Eq. (\ref{Mises}) this implies that on
the crack interface
\begin{equation}
\sigma_{1} =  \sqrt{3}~\sigma_{_{\rm Y}}\!,~~~~~~\sigma_{2} = 0.
\label{PS}
\end{equation}

The typical scale $R$ follows from the physics of the nucleation
process. It is physically plausible that void formation is more
susceptible to hydrostatic tension than to distortional stresses. We
assume that void nucleation occurs where the hydrostatic tension
$P$,  $P\equiv\case{1}{2}$Tr$\B \sigma$, exceeds some threshold
value $P_c$. The calculated the hydrostatic tension $P$ along the
yield curve is found to be significantly higher than its value on
the crack surface $P=\case{\sqrt{3}}{2}\sigma_{_{\rm Y}}$ (cf. Eq.
(\ref{PS})). On the other hand the linear-elastic solution implies a
monotonically decreasing P outside the yield curve. We thus expect P
to attain its maximum value between the blunted crack tip and the
yield curve. This conclusion is fully supported by finite elements
method calculations, cf. \cite{77McM}. As the detailed
elastic-plastic crack tip fields are not computed here, we use the
outer elastic solution on the yield curve to determine the void
nucleation position. We expect that this approximation should not
affect the roughening exponents on large scales. The distance from
the crack tip where $P$ attains its maximal value is of the order of
the size of the plastic zone.

In this model the nucleation occurs when P exceeds a threshold
$P_c$. The void will thus appear at a typical distance $R$. A very
clear demonstration of the appearance of the void near the boundary
of the plastic zone is seen in the molecular dynamics simulations of
\cite{99Falk}. Note that $R$ is not a newly found length scale; it
is the well known scale of the plastic zone \cite{68Rice}. Its
identification with a length scale that is related to the scaling
behavior of fracture surfaces is however new. This stems from the
proposition that positive correlations appear only between the
positions of nucleated voids. Below $R$ one enters the regime of
plastic processes whose theory is far from being settled. We should
also comment that it is possible that positive correlations appear
even below the scale of the plastic zone since experiments indicate
that several voids nucleate within the plastic zone
\cite{99BP,03Cil}. Finally, we note that the assumption of perfect
plasticity, i.e. that $\sigma_{_{\rm Y}}$ is independent of the
state of deformation and its history, is {\em not} true for real
materials; usually $\sigma_{_{\rm Y}}$ is not sharply defined; it
can increase with plastic deformations \cite{90Lub}. This
phenomenon, known as ``work-hardening'' or ``strain-hardening'' is
not taken into account in this simple model, with the hope to be
irrelevant for the crack morphology on large scales.

Naturally, the precise location of the nucleating void will
experience a high degree of stochasticity due to material disorder.
A basic assumption behind our treatment of this stochasticity is
that the material disorder occurs on scales smaller than $R$,
allowing us to introduce a probability distribution function that is
a function of the hydrostatic tension on the yield curve. Since we
do not know from first principles the probability distribution for
void formation, we consider in our model below two possible
distribution functions. In all cases nucleation cannot occur if
$P<P_c$. For
$P>P_c$ the void occurs with probability\\
\begin{eqnarray}
P&\propto& P-P_c\ , \label{Plin}\\
P&\propto& \exp[\alpha(P-P_c)]-1 \ . \label{Pexp}
\end{eqnarray}
In the exponential case we considered two different values of
$\alpha$. In Fig. \ref{straightsteps} we show three such pdf's as
they appear for a perfectly straight crack. We note that these
distributions are symmetric about the forward direction.
Nevertheless they have sufficient width to allow deviations from
forward growth. These deviations will be responsible later for the
roughening of the crack. For comparison examine also the pdf's for a
general crack which are shown in Fig. \ref{pdf}. There the symmetry
is lost: correlation to previous steps create a preference for the
upward direction.

\begin{figure}[here]
\centering \epsfxsize=  7 truecm \epsfbox{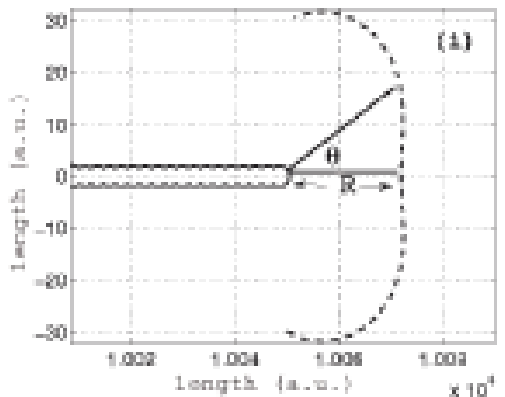}
\epsfxsize= 7 truecm \epsfbox{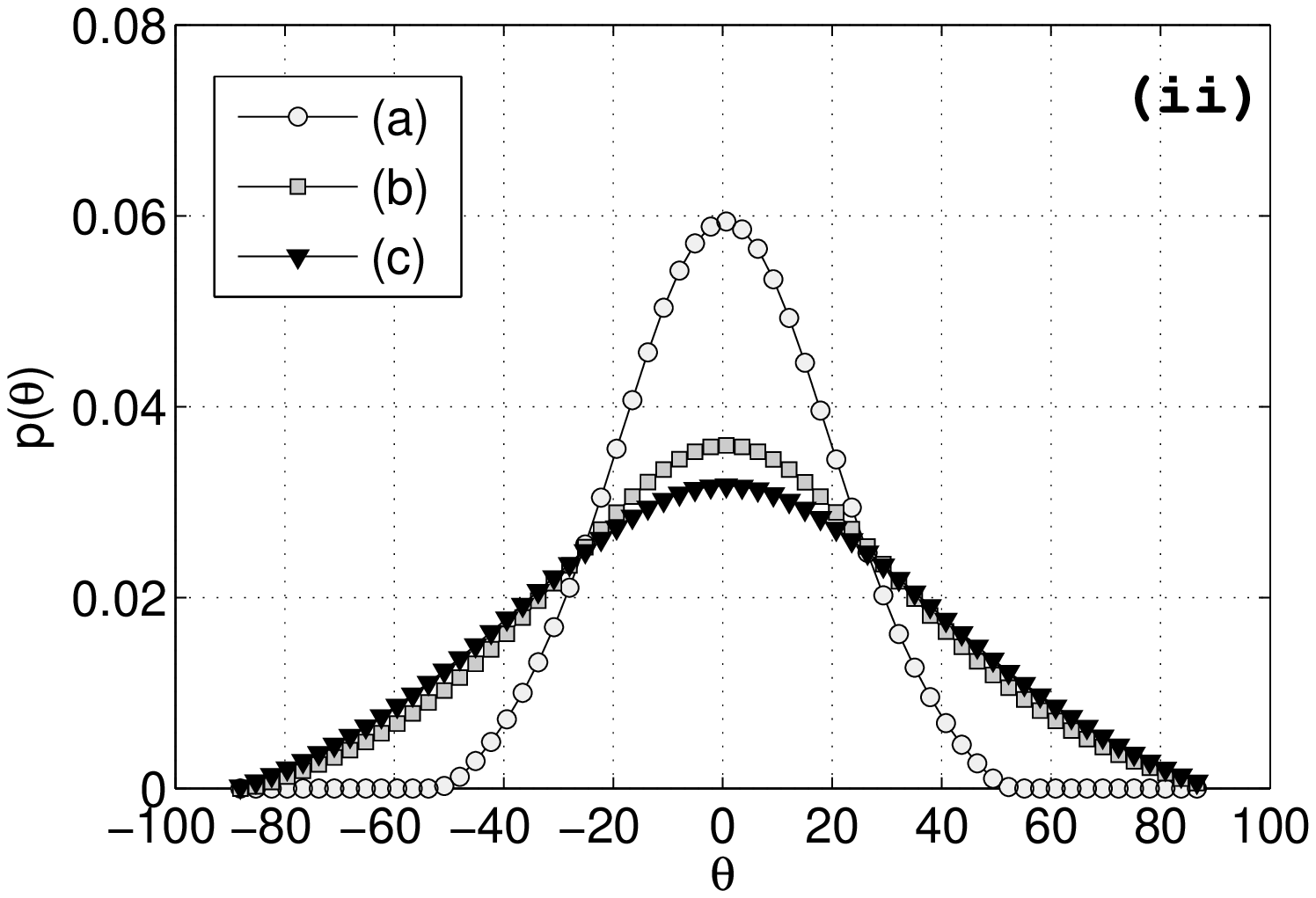}
 \caption{Panel (i): the
tip of a straight crack and the yield curve in front of it. \\ Panel
(ii): three probability distribution functions calculated for the
configuration in (i). The abscissa is $\theta$, the angle measured
from the crack tip as seen in panel (i). The ordinate is the
normalized probability (per unit $\theta$) to grow in the $\theta$
direction. The distributions are symmetric and wide enough to allow
deviations from the forward direction. For all the curves
$\frac{\sigma_{_{\rm Y}}}{\sigma^\infty} = 6$. For curve (a)
$\mathrm{p}(\theta) \propto \exp[(P-P_c)]-1$ and
$\frac{P_c}{\sigma^\infty}=8$, for curve (b) $\mathrm{p}(\theta)
\propto \exp[0.2 (P-P_c)]-1$ and $\frac{P_c}{\sigma^\infty}=6$ and
for curve (c) $\mathrm{p}(\theta) \propto P-P_c$ and
$\frac{P_c}{\sigma^\infty}=6$.} \label{straightsteps}
\end{figure}

Each growth step in our model is composed of two events. Firstly the
material yields near the crack tip, creating a plastic zone with a
void growing somewhere at the zone boundary. Secondly the crack tip
and the void coalesce. We note that there is a separation of time
scales between these two events. The first is slow enough to be
governed by a quasi-static stress field. The second event occurs on
a shorter time scale. It is clear that we forsake in the model any
detailed description of the geometry on scales smaller than $R$. Any
relevant scaling exponent that will be found in this model will
refer to roughening on length scales larger than $R$. Therefore, the
physical process in which the crack coalesces with the multiple
voids ahead of it is substituted by a single void coalescence with
the crack.

\subsection{Results}

Calculating the stress field around the crack, according to the
method of iterated conformal maps, we can readily find the yield
curve and the physical region in its vicinity where a void can be
nucleated. Choosing with any one of the probability distributions
described above, we use this site as a pointer that directs the
crack tip. We then use the method of iterated conformal mappings to
make a growth step to coalesce the tip with the void. Naturally the
step sizes are of the order of $R$. We reiterate that this model
forsakes the details of the void structure and all the length scales
below $R$. Since we are making {\em linear} steps below $R$, we
anticipate having an artificial scaling exponent $H=1$ for scales
smaller than $R$. This is clearly acceptable as long as we are
mainly interested in the scaling properties on scales larger than
$R$.

\begin{figure}[top]
\centering \epsfig{width=.48\textwidth,file=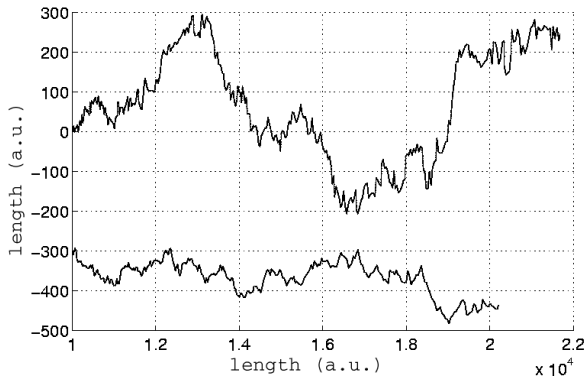}
\caption{Two typical cracks generated with our model. Note the huge
difference in scales of the abscissa and ordinate (the cracks are
globally flat) and that the lower crack had been translated by -300.
The upper crack exhibits two decades of self-affine scaling with a
Hurst exponent $0.64$. The lower crack has smaller standard
deviation and therefore a shorter scaling range. Nevertheless it
appears that in its shorter scaling range it exhibits an exponent
that is very close to the upper crack.} \label{rough crack}
\end{figure}

In Fig. \ref{rough crack} we present two typical cracks that were
grown using this method.  Both cracks were initiated from a straight
crack of length 10000, representing the experimental paradigm of
introducing a notch before loading the sample. The upper crack was
grown using the broader exponential pdf of Fig. \ref{pdf} curve (b).
The lower crack was grown with the narrower pdf of Fig. \ref{pdf}
curve (a). Clearly, the upper crack exhibits stronger height
fluctuations, as can be expected from the wider pdf and the choice
of parameters. For the lower crack forward growth is much more
preferred. In the upper crack the positive correlations between
successive void nucleation and coalescence events can be seen even
with the naked eye. This is precisely the property that we were
after. A neat way to see this tendency is in the pdf's as they are
computed on the yields crack for a typical, rather than straight,
crack. In Fig. \ref{pdf} we show these pdf's for the crack whose
yield curve is shown in the upper panel. We see that now the
symmetry of the pdf's is lost, and positive values of $\theta$ are
preferred. This is the source of positive correlations that
eventually gives rise to a non-trivial roughening exponent.

\begin{figure}
\centering \epsfxsize=  7 truecm \epsfbox{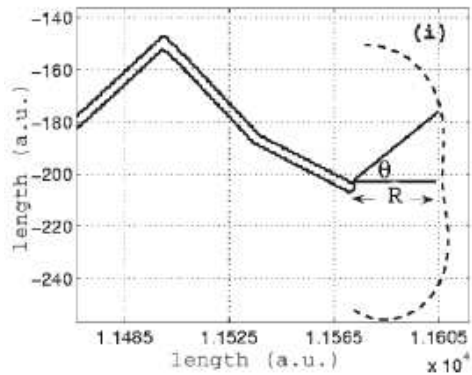}
\epsfxsize= 7 truecm \epsfbox{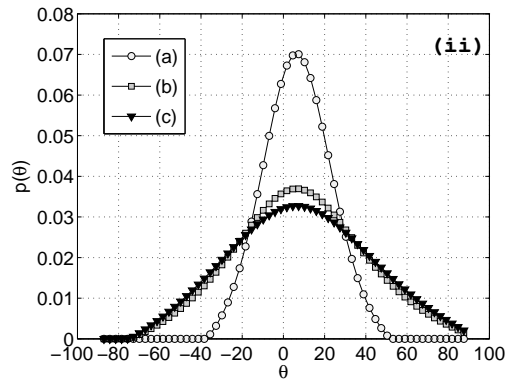} \caption{Panel (i):
the tip of a ``rough'' crack and the yield curve in front of it.
Panel (ii): three probability distribution functions calculated
for the configuration in panel (i). The abscissa is $\theta$, the
angle measured from the crack tip as seen in panel (i). The
ordinate is the normalized probability (per unit $\theta$) to grow
in the $\theta$ direction. The pdf's are those used in Fig.
\ref{straightsteps}, using the same parameters. Note the upward
preference in all the pdf's due to the broken symmetry.}
\label{pdf}
\end{figure}
This is born out by the measurements of the scaling properties of
the fracture lines morphology that we discuss next.

We first restrict ourselves to a monoscaling analysis. Due to the
significant computational cost of the iterated conformal maps
technique the numerical investigation of the growth model had a
limited number of realizations of a few hundreds growth steps. As a
result of the relative paucity of data, the structure functions
defined in Eq. (\ref{Sn}) would not converge well enough to provide
reliable exponents. Comparing the various available methods for
estimating roughness exponents \cite{95SVR} we decided to select the
max-min method, which seems to give reliable results in our case.
Therefore, one defines $S(\ell)$ according to
\begin{equation}
S(\ell)\equiv\left<Max\left\{y(\tilde x)\right\}_{x<\tilde
x<x+\ell}- Min\left\{y(\tilde x)\right\}_{x<\tilde
x<x+\ell}\right>_x \ . \label{rough}
\end{equation}
For self-affine graphs the Hurst exponent $H$, as was explained in
the introduction, is obtained via the scaling relation
\begin{equation}
S(\ell) \sim \ell^{H} \ . \label{Hurst}
\end{equation}

In Fig. \ref{h(r)} we present a typical log-log plot of $S(\ell)$
vs. $\ell$, in this case for the two cracks in Fig. \ref{rough
crack} with power-law fits of $H=0.64$ and $H=0.68$ respectively.
Indeed, as anticipated from the visual observation of Fig.
\ref{rough crack} the exponent is higher than 0.5.

\begin{figure}[here]
\centering \epsfig{width=.49\textwidth,file=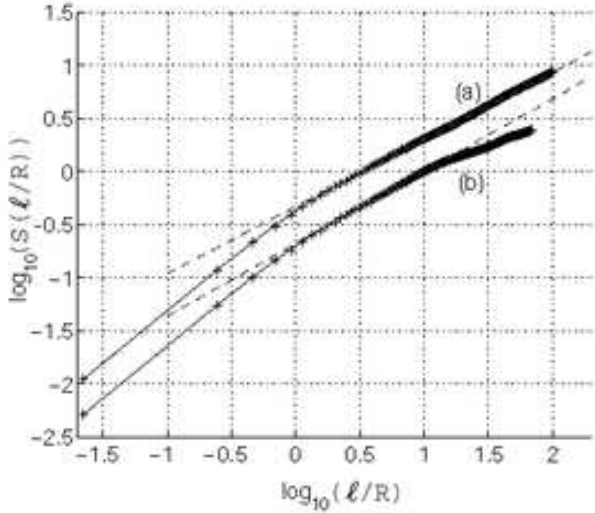}
\caption{Calculation of the roughening exponent $H$. The slopes of
the dotted lines are $0.64$ for the upper plot (curve a) and 0.68
for the lower (curve b). Note that the initial scaling with slope
$1$ is relevant for length scales smaller than $R$. This scaling
is an artifact, resulting from our algorithm that connects the
crack tip to a void by a straight line.} \label{h(r)}
\end{figure}
It turned out that all the cracks grown by our algorithm gave rise
to scaling plots in which a scaling range with $H=0.66\pm 0.03$ is
clearly seen. When the pdf allowed for a sizeable standard
deviation, the cracks gave a very nice scaling plot with at least
two decades of clear anomalous scaling. When the standard deviation
was small, the scaling range was more meager, as seen in Fig.
\ref{h(r)}. It is interesting to stress that the anomalous scaling
exponent appears insensitive to the pdf used, although the extent of
the scaling range clearly depended on the pdf. We note that our
measured scaling exponents are very close to the exponents observed
in other two-dimensional experiments \cite{92PABT, 93KHW, 94EMHR,
03SAN}. In addition the value of $R$ does not effect the scaling
properties of a crack, i.e. it doesn't seem to matter how long the
step is, so long as a wide distribution of angles is allowed.

We now turn to a multiscaling analysis of the morphology of fracture
lines. For this purpose, we define
\begin{equation}
\tilde{S}_n(\ell)\equiv \langle |Max\left\{h(\tilde
x)\right\}_{x<\tilde x<x+\ell}- Min\left\{h(\tilde
x)\right\}_{x<\tilde x<x+\ell}|^n\rangle \ . \label{newS}
\end{equation}
 The scaling exponents $\zeta^{(n)}$ are defined in
analogous way to Eq. (\ref{zetan}) by
\begin{equation}
\tilde{S}_n(\lambda \ell)\sim \lambda^{\zeta^{(n)}}
\tilde{S}_n(\ell) \ .
\end{equation}
Note that $\tilde{S}_1(\ell)$ is just $S(\ell)$ defined in Eq.
(\ref{rough}) and $\zeta^{(1)}$ is just $H$ defined in Eq.
(\ref{Hurst}).

\begin{figure}[here]
\centering \epsfig{width=.45\textwidth,file=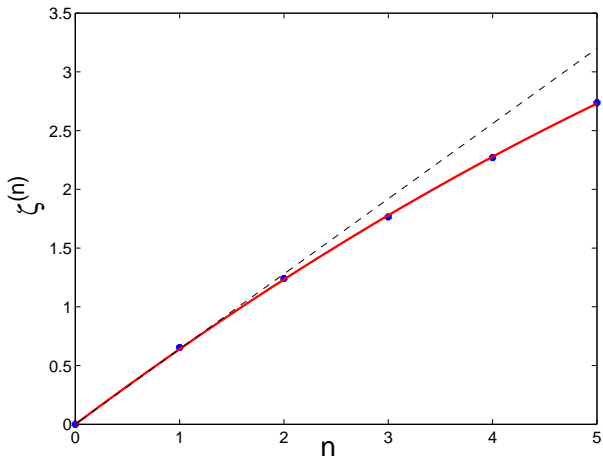}
\caption{The spectrum $\zeta^{(n)}$ as a function of the moment
order $n$ for rupture lines in the model of \cite{04BMPb, 05ABKMP}.
The function is fitted to the form $\zeta^{(n)} = nH-n^2\lambda$ and
the parameters $H$ and $\lambda$ are given. The errors in the
estimation of these parameters reflect both the variance between
different realizations and fit quality. The linear plot
$n\zeta^{(1)}$ is added for comparison.}\label{model}
\end{figure}

The resulting exponents $\zeta^{(n)}$ for the cracks generated by
this model are shown in Fig. \ref{model}. For the low orders
moments (here we are limited by the paucity of data to $n\le 5$)
one again fits  a quadratic function, with $H = 0.66 \pm 0.03$ and
$\lambda = 0.023 \pm 0.003$. The errors in the estimation of these
parameters reflect both the variance between different
realizations and the fit quality. The $n$ dependence of the
exponents $\zeta^{(n)}$ and the values of the fitting parameters
are in agreement with the experimental ones. Since there is
nothing in the model that is specific for the physics of paper, it
appears that multiscaling is a generic property of the fracture
process, at least in 1+1 dimensions.

To sum up, we have developed a model that shows that long-range
elastic interactions, the appearance of a finite length scale in
the growth process and material disorder are able to reproduce the
scaling properties associated with the morphology of 1+1
dimensional fracture. It should be stressed that a deeper
understanding of the nature of the long-range correlations induced
by geometrical irregularities of the crack  is still lacking. It
would be nice to specify which physical quantity controls the
growth at a finite length scale $R$ away from the crack tip.
Moreover, it is important to study this model with different local
crack tip physics to better understand the degree of universality
and the role of the length scale $R$.

\section{Scaling Properties in 2+1 dimensional Fracture}
\label{SO2}

In previous sections we have considered the scaling properties of
1+1 dimensional rupture lines. In this section we extend the
discussion to include fracture surfaces which are 2+1 dimensional
graphs. In 2+1 dimensions one denotes the graph as $h(\B r)$ and considers
again the structure function $S_2(\B \ell)$, which is now a two
dimensional function
\begin{equation}
 S_2(\B \ell)\equiv \langle (h(\B r+\B \ell)-h(\B r))^2\rangle \ , \label{S2}
 \end{equation}
where angular brackets denote an average over all $\B r$. Initially
no attention was paid to the fact that the isotropy in the fracture
plane is broken due to initial conditions that lead to a preferred
propagation direction and the statement of \cite{84MPP} was that the
structure function is a homogeneous function of its arguments,
$S_2(\lambda \B \ell)\sim \lambda^{\zeta^{(2)}} S_2(\B \ell)$, as
implied by Eq. ({\ref{def_affine}). In fact such a statement is
tenable only if the fracture process {\em and} the material are
isotropic. Usually the crack propagates predominantly in one
direction (say  $\B {\hat x}$) and the vector $\B \ell$ defines an
angle $\theta$ with respect to  $\B {\hat x}$, $\theta =\cos^{-1}
(\B {\hat x}\cdot \B {\hat \ell})$. There is no reason why the
scaling exponent $\zeta^{(2)}$, if it exists at all, should not
depend on the angle $\theta$. Indeed, in the later work that
followed \cite{84MPP} this problem was recognized and scaling
exponents were sought for one dimensional cuts through $S_2(\B
\ell)$, typically parallel and orthogonal to the direction of the
crack propagation. Besides the obvious meaning of `parallel' and
`orthogonal' to $\B {\hat x}$, no reason was ever given why these
particular directions are expected to provide clean scaling
properties. We will argue below that in general such one dimensional
cuts exhibit a mixture of scaling exponents with amplitudes that
depend on the angle $\theta$, where $\theta=0$ and $\theta=\pi/2$
are not special. To better understand the way crack propagation
directionality affects scaling isotropy, imagine a fracture
experiment in which an initial circular cavity is made to propagate
by a tensile load such that the crack edge remains circular on the
average, without any preferred propagation direction in the plane
normal to the load, see Fig. \ref{experiment}. From the point of
view of the scaling properties of the rough fracture surface that is
left behind the advancing crack, such an experiment is the analogue
of homogenous and isotropic turbulence in nonlinear fluid mechanics.
Normal experiments in both fracture and turbulence involve symmetry
breaking; the boundary conditions introduce anisotropy, making the
discussion of scaling properties non-trivial. In turbulence it was
shown how to disentangle the anisotropic contributions from the
isotropic one by projecting the measured correlation and structure
functions on the irreducible representations of the SO(3) symmetry
group \cite{05BP}. The scaling phenomena seen in the isotropic
sector of anisotropic experiments are identical to those expected in
the hypothetical experiment of homogenous and isotropic turbulence.
In this section we describe how a similar concept is introduced to
the field of fracture: we will show that decomposing the
height-height structure functions of fracture surfaces into the
irreducible representations of the SO(2) symmetry group results in a
simplification  and rationalization of the scaling properties that
is not totally dissimilar to the one obtained in turbulence. The
scaling properties of the isotropic sector should
 be observable in principle in an experiment like the one
shown in Fig. \ref{experiment}, which contrary to turbulence may
be performed in reality.

 \begin{figure}
\begin{center}
\epsfysize=4.0 truecm \epsfbox{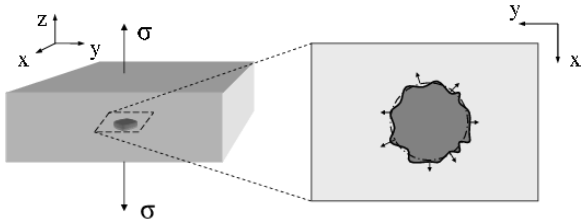} \caption{Sketch of a
hypothetical fracture experiments arranged to allow a crack to
develop in an isotropic fashion, i.e. with all angles $\theta$ being
statistically equivalent.On the left, the full three dimensional
experiment is shown, where the tensile axis is along $z$ and a
circular cavity is in the $xy$ plane. On the right, a magnified
version of the circular cavity in the $xy$ plane is shown.}
\label{experiment}
\end{center}
\end{figure}

Given an experimental surface $h(\B r)$ we first compute the
second order structure function Eq. (\ref{S2}). The vector $\B
\ell$ is associated with a norm $\ell$ and an angle $\theta$. By
construction, the second order structure function is symmetric
under $\theta \to \theta+\pi$.
 Accordingly, decomposing the structure functions into the irreducible representations of the SO(2)
 symmetry group results in summations over even indices only:
 \begin{equation}
 S_2(\ell , \theta) = \sum_{m =\; -\infty}^{\infty} a_{2m}(\ell)
e^{i 2 m \theta} \ . \label{decompose}
\end{equation}
Such a decomposition is deemed useful when each of the scalar
functions $a_{2m}(\ell)$ is itself a homogeneous function of its
argument, characterized by an $m$ dependent exponent:
    \begin{equation}
        |a_{2m} (\lambda \ell )| \sim \lambda^{\zeta^{(2)}_{2m} }|a_{2m}(\ell)| \ ,
    \end{equation}
where $|\cdot |$ stands for the norm of a complex number. For an
isotropic fracture in an isotropic medium we expect $a_{2m}(\ell)=0$
for all $m\ne 0$. In usual mode I experiments in which the crack
propagates along the $\B {\hat x}$ direction and the tensile load is
in the normal direction, there should be the same physics along
lines with angles $\theta$ and $-\theta$. This invariance under
$\theta\to -\theta$ implies that the arguments of all
$a_{2m}(\ell)\ne 0$ should be 0 or $\pi$. In reality this invariance
might not hold on large length scales due to the paucity of data or
due to some symmetry breaking process, see below.

Our first experimental example was obtained \cite{98AB} from a
compact tension specimen made of 7475 aluminum alloy first
precracked in fatigue and then broken under tension in mode I. The
raw fracture surface and the second order structure function
computed from it are shown in Figs. \ref{raw} and \ref{FigS2}
respectively.
 \begin{figure}
\begin{center}
\epsfysize=5. truecm \epsfbox{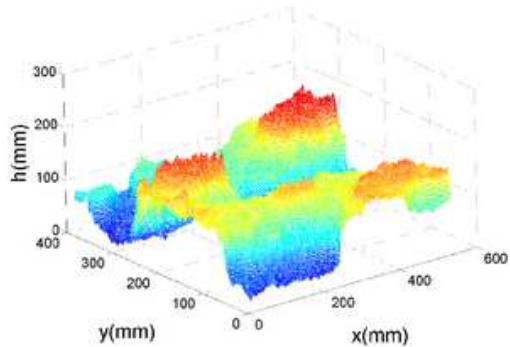} \caption{The raw
fracture surface of the aluminum alloy obtained in Ref.
\cite{98AB}.}
 \label{raw}
\end{center}
\end{figure}

\begin{figure}
\begin{center}
\epsfysize=5 truecm \epsfbox{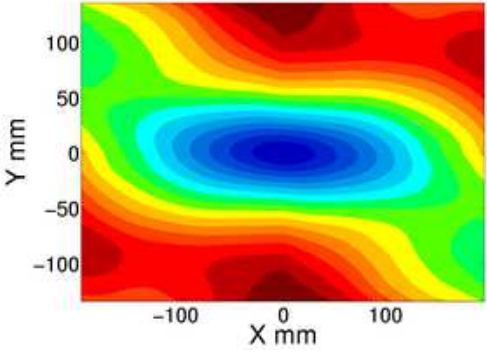} \caption{Contour plot of
the second order structure function of the surface shown in Fig.
\ref{raw}.}
 \label{FigS2}
\end{center}
\end{figure}
One sees the anisotropy of $S_2(\ell)$ with the naked eye. To
quantitatively characterize this anisotropy, the structure
function was decomposed as in Eq. (\ref{decompose}). The log-log
plots of $a_0(\ell)$, $2|a_2(\ell)|$ and $2|a_4(\ell)|$ are
exhibited in Fig. \ref{alua4}.

 \begin{figure}
\begin{center}
\epsfxsize=6.5 truecm \epsfbox{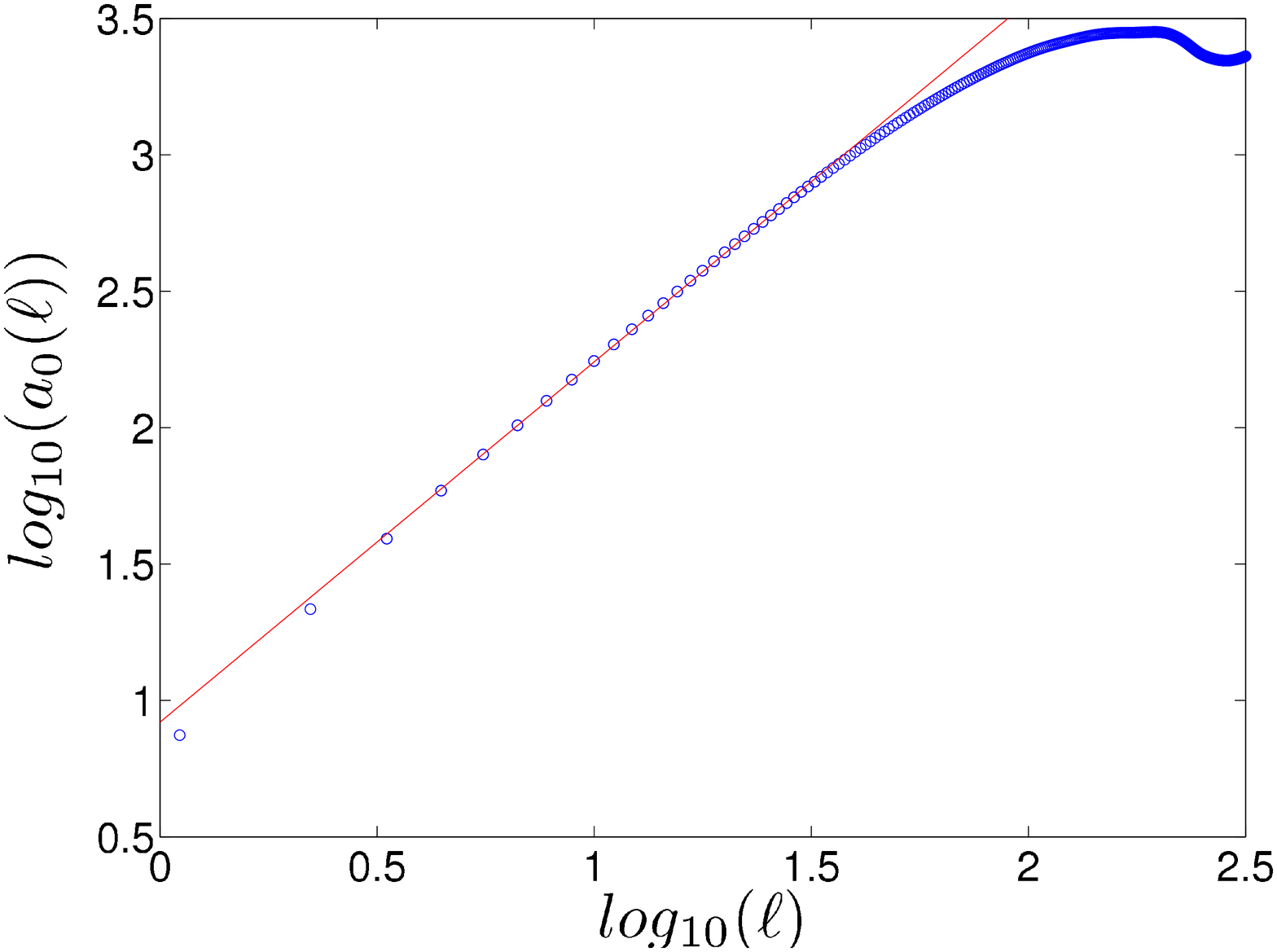} \epsfxsize=6.5 truecm
\epsfbox{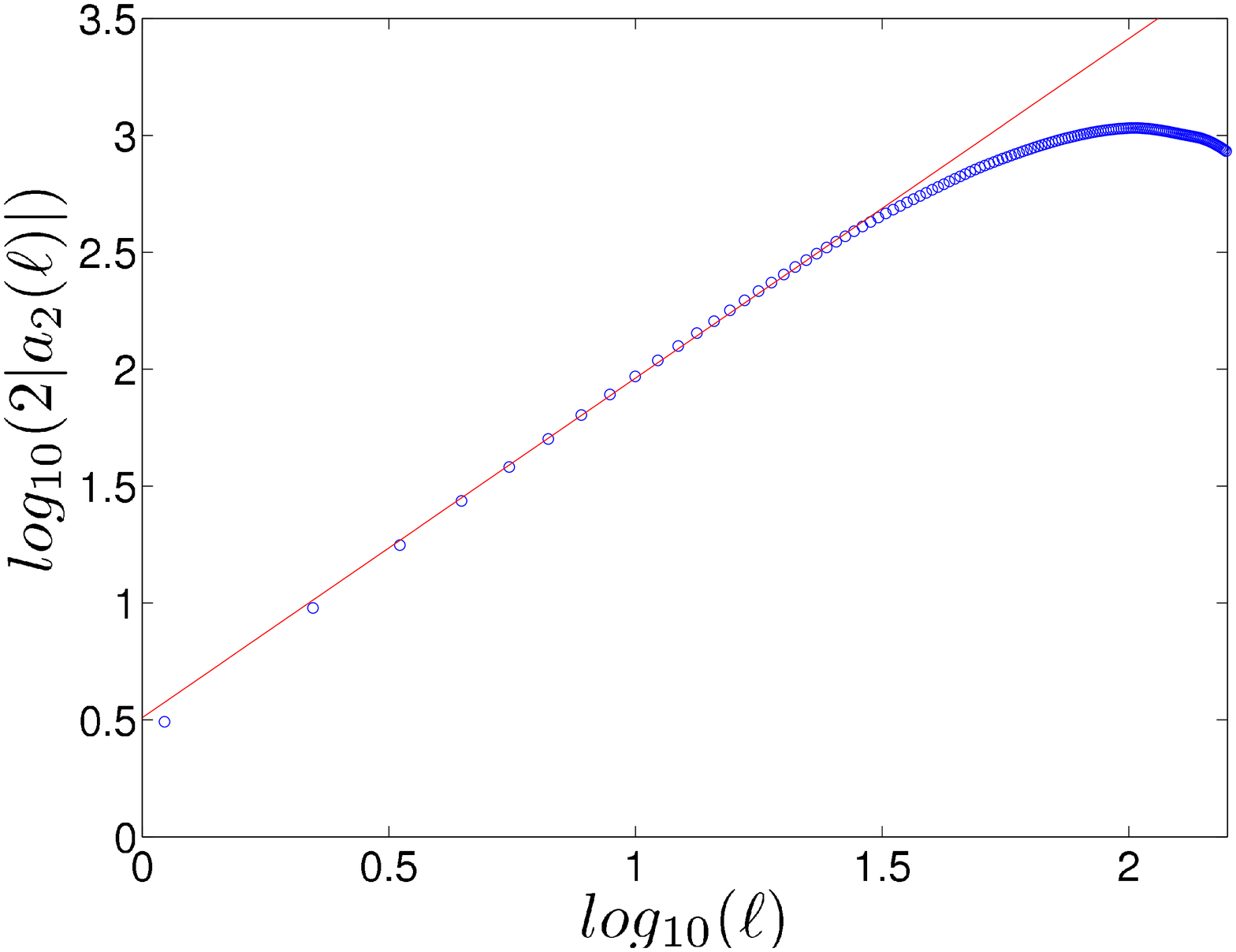} \epsfxsize=6.5 truecm
\epsfbox{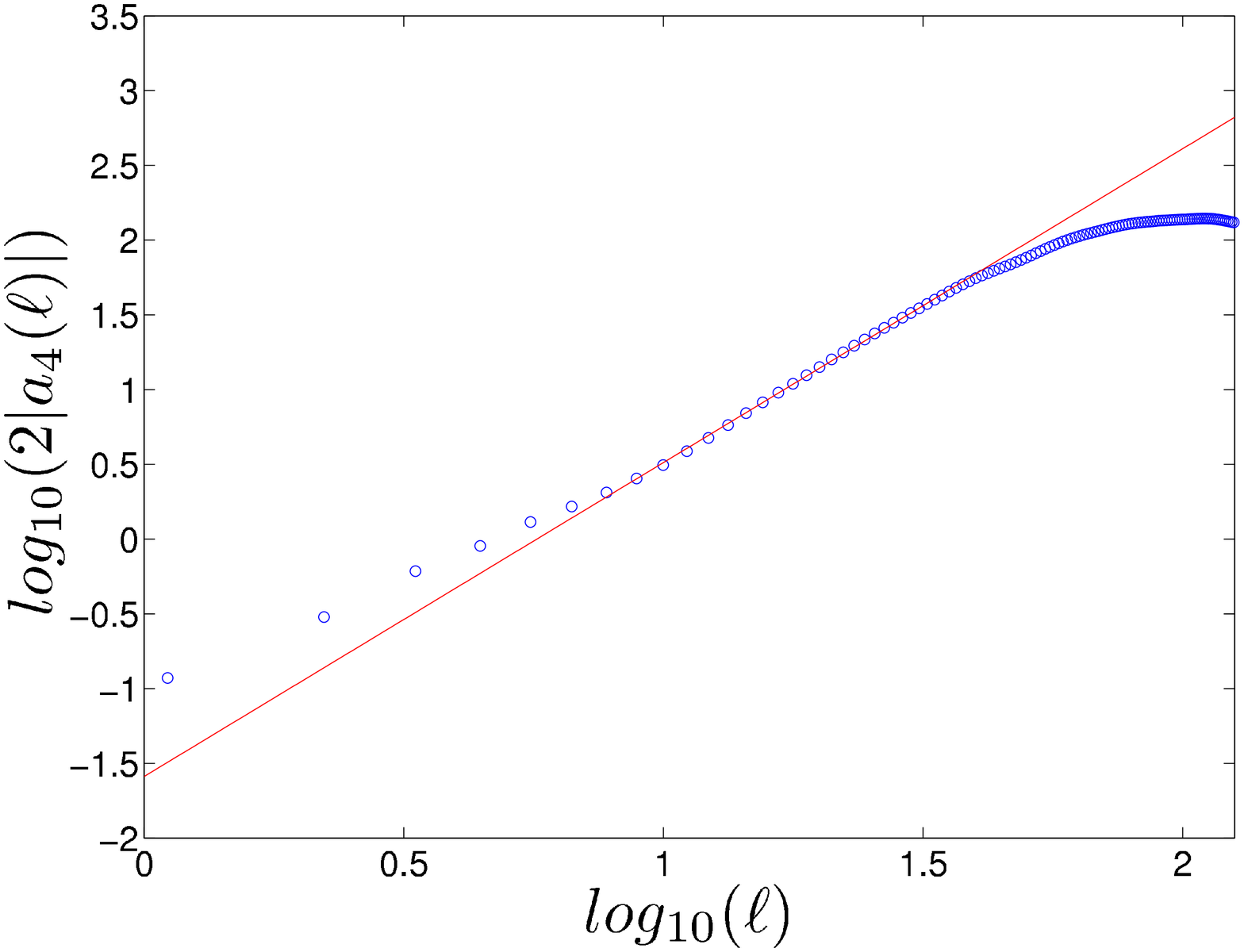} \caption{Log-log plots of the amplitudes
$a_0(\ell)$, $2|a_2(\ell)|$ and $2|a_4(\ell)|$ vs. $\ell$ for the
aluminum alloy.}
 \label{alua4}
\end{center}
\end{figure}

By performing linear fit of the relevant range in the log-log
plots we find the following exponents
\begin{equation}
\zeta^{(2)}_0 \!\!= 1.32 \pm 0.07 \  ,  \zeta^{(2)}_2\!\! = 1.45
\pm 0.08 \ , \zeta^{(2)}_4 \!\!= 2.1\pm 0.1\ . \label{metzeta}
\end{equation}
The implication is that at smaller length-scales the smaller
exponent $\zeta^{(2)}_0$ should be dominant and vice versa.
Indeed, examining again the contour plot in Fig. \ref{FigS2} one
observes that at small scales the contours tend to ellipses of
smaller eccentricity, whereas at larger scale the contours are
ellipses with increasing eccentricity.

The crucial test of this approach is whether one can reconstruct
the structure function $S_2(\ell,\theta)$ using the functional
form of the irreducible representation and a minimal number of
parameters. Indeed, at smaller values of $\ell$ the first two
irreducible representations suffice. Writing
\begin{equation}
S_2( \ell, \theta) \approx 8.30~ \ell^{1.32}  +3.22 ~ \ell^{1.45}
cos(2\theta + \pi) \ , \label{represent}
\end{equation}
we compare in Fig.\ref{compare} the experimental data to Eq.
(\ref{represent}) for $\ell=5$, 15 and 25mm. The excellent fit is
obvious. In fact, with four parameters (two amplitudes and two
exponents) we can represent the structure function to within 1\%
in $L^2$ norm as long as $\ell\le 30mm$. For larger values of
$\ell$ the agreement decreases, and we need to employ the next
irreducible representation. Adding
$0.026~\ell^{2.1}\cos(4\theta+\pi)$, we find the fit shown in Fig.
\ref{35} for $\ell=35mm$. Beyond these values the power-laws fits
lose their credibility for this experimental data set.

\begin{figure}
\epsfxsize=6 truecm \epsfbox{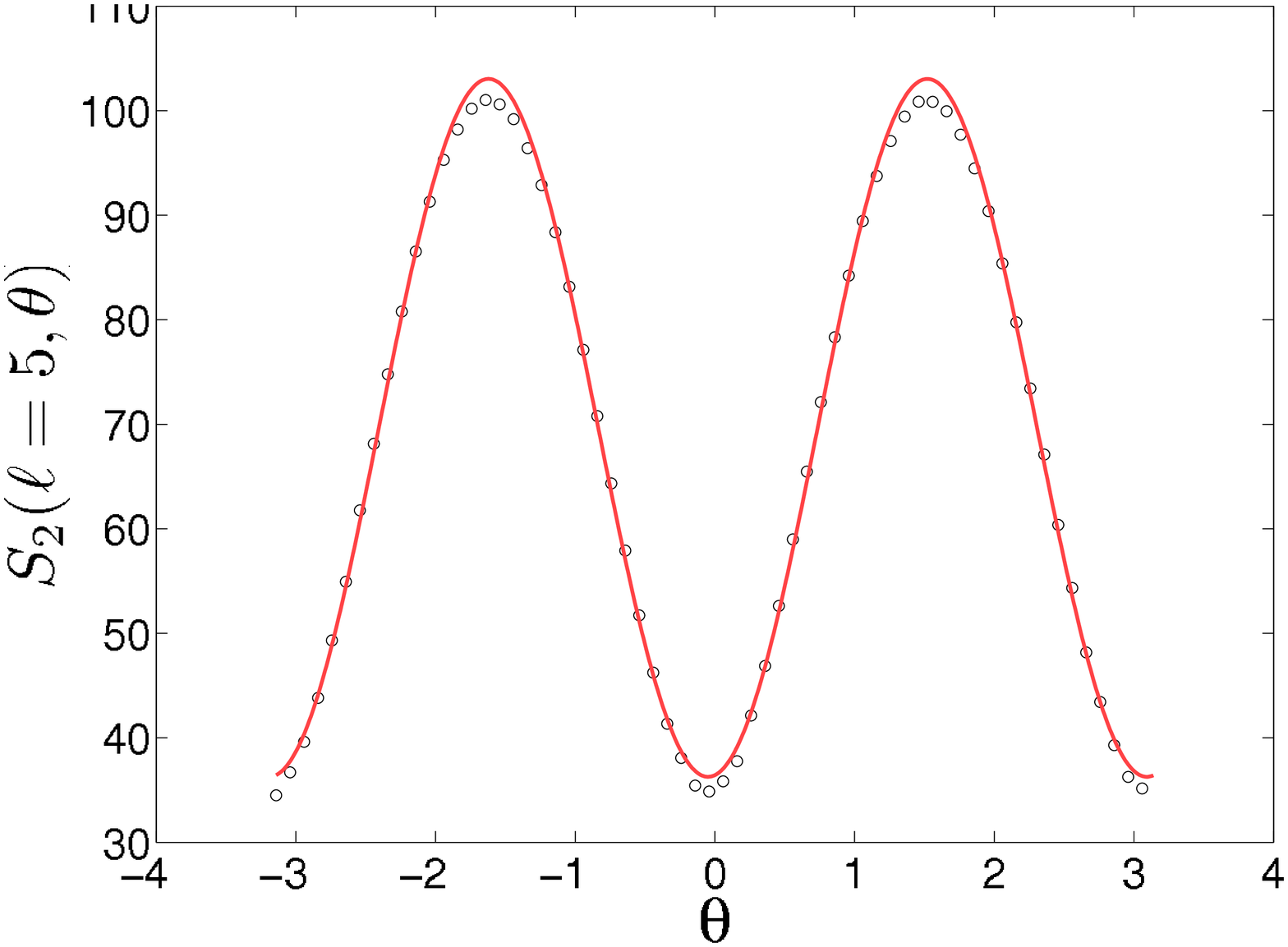} \epsfxsize=6 truecm
\epsfbox{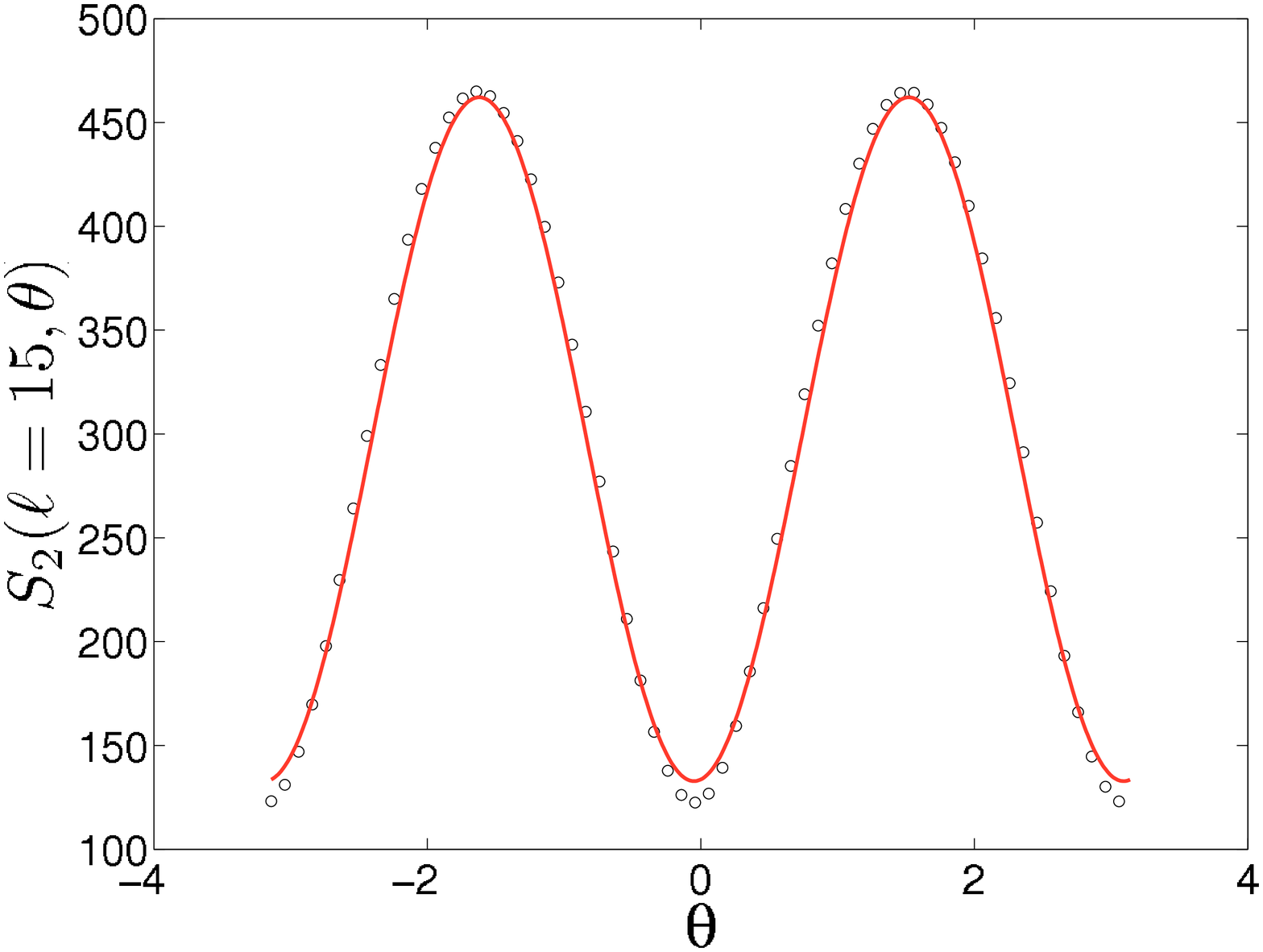} \epsfxsize=6
truecm \epsfbox{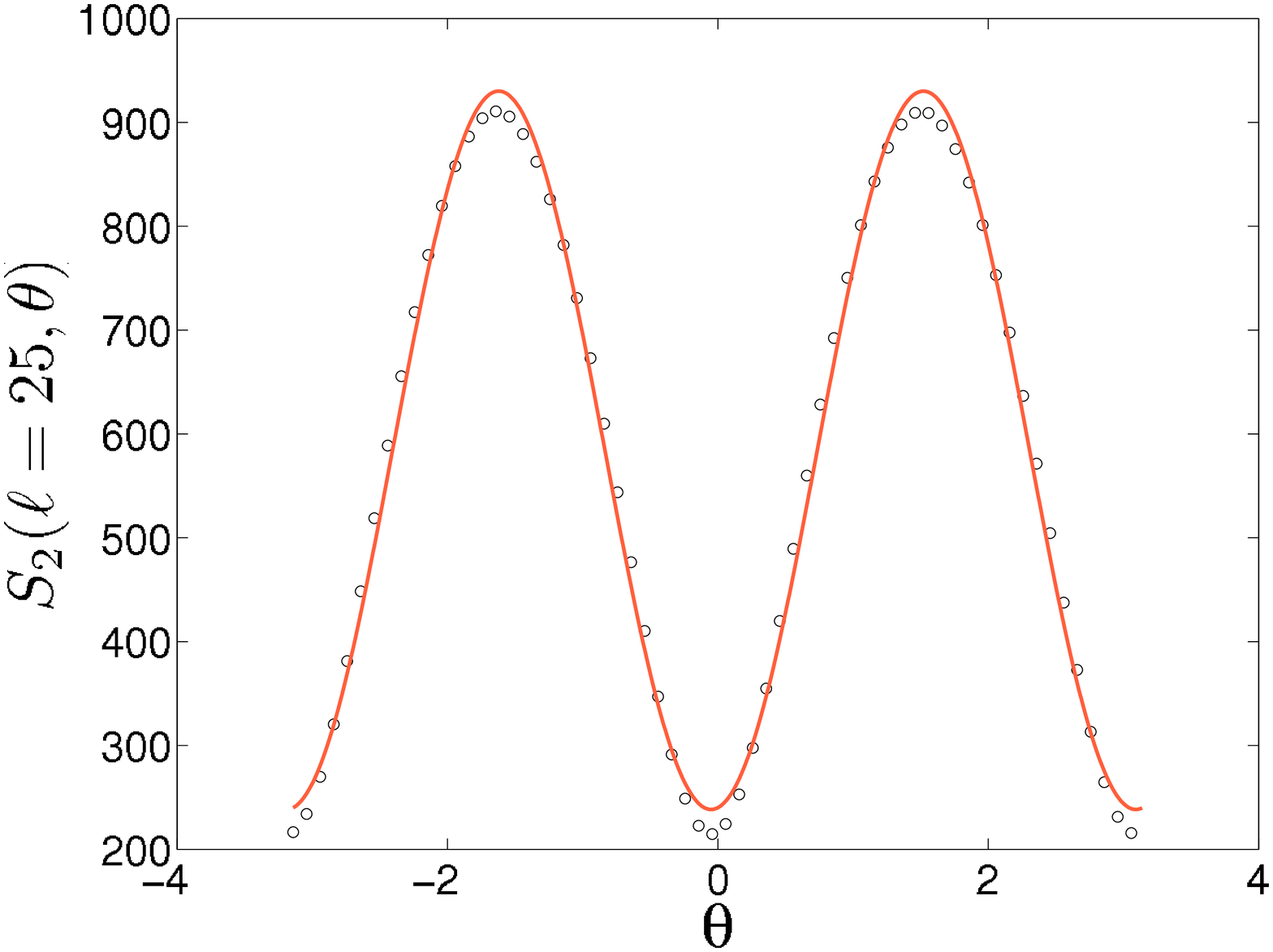} \caption{The experimental $S_2(\ell,
\theta)$ for the aluminum alloy (circles) and the representation Eq.
(\ref{represent}) (line), for $\ell=5$, 15 and 25mm.}\label{compare}
\end{figure}

\begin{figure}
\epsfxsize=6 truecm \epsfbox{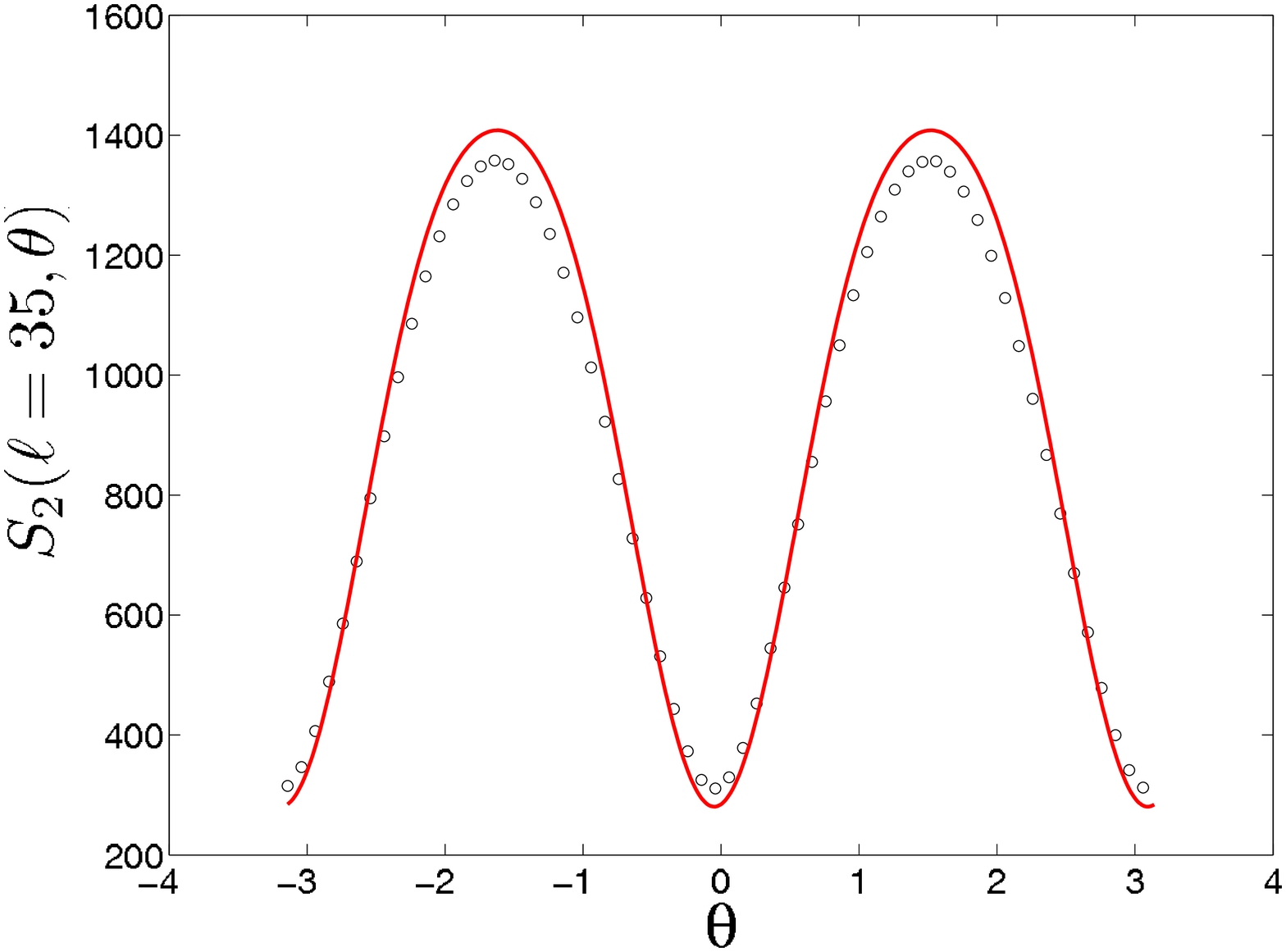} \caption{The
experimental $S_2(\ell, \theta)$ for the aluminum alloy (circles)
and the SO(2) expansion up to the third even order irreducible
representation (line), for $\ell=35mm$.} \label{35}
\end{figure}

A second experimental example was obtained from the dynamic fracture
of artificial rocks produced from carbonatic aggregates cemented by
epoxy \cite{05Sag}. The samples are plates of size $400 \times
400\times 9$ mm and the fracture surface was measured using a
scanning laser profilometer. The analysis of the experimental data
follows verbatim the first example. The plots of $a_0(\ell)$ and
$2|a_2(\ell)|$ are shown in Fig. \ref{a0a2rock}.

\begin{figure}
\epsfxsize=6 truecm \epsfbox{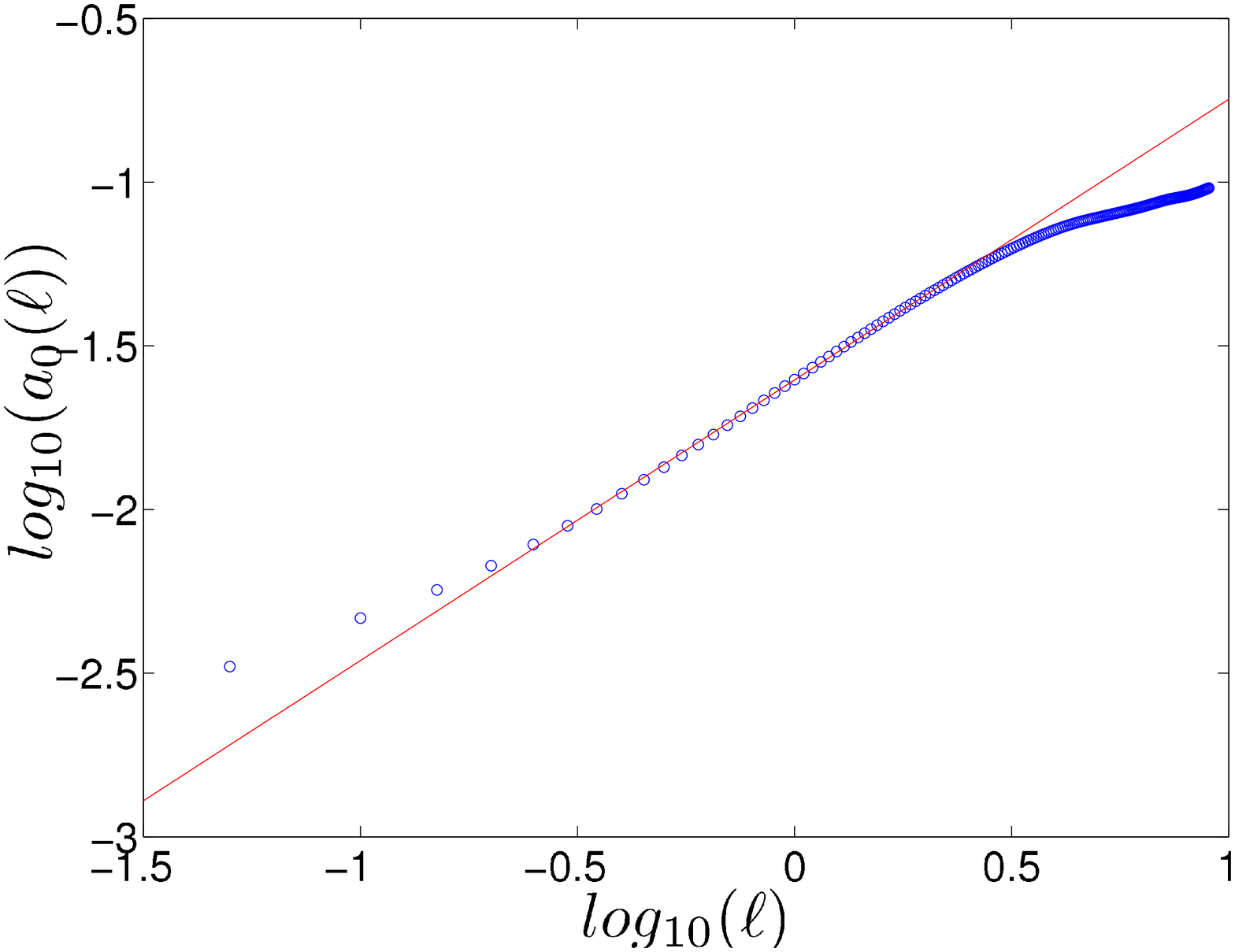}
\epsfxsize=6truecm\epsfbox{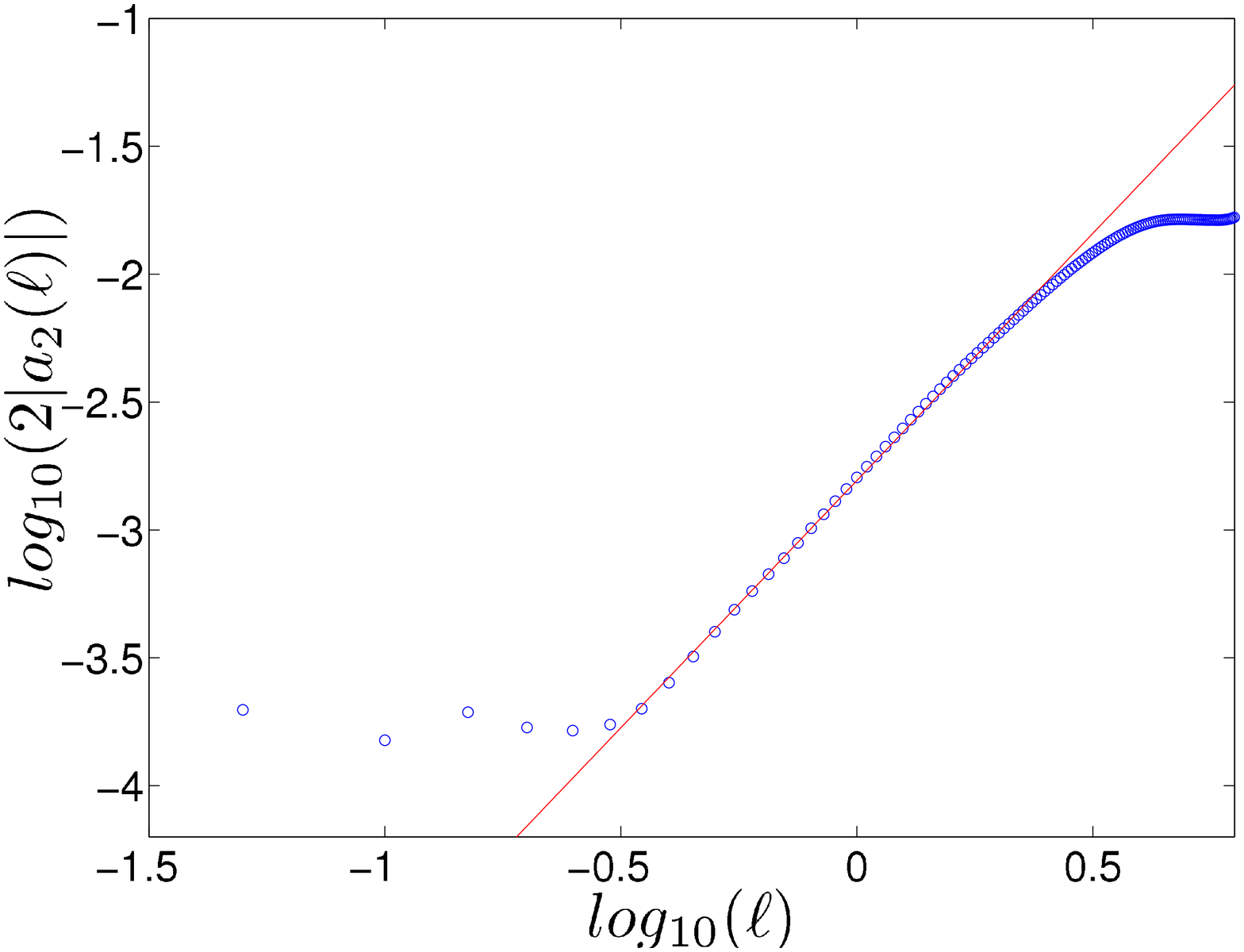} \caption{Log-log plots of
the amplitudes $a_0(\ell)$ and $2|a_2(\ell)|$ for the artificial
rock.} \label{a0a2rock}
\end{figure}
Fitting the plots we find
\begin{equation}
\zeta^{(2)}_0 \!\!= 0.86 \pm 0.05\ , \zeta^{(2)}_2\!\! = 1.93 \pm
0.05\ , \zeta^{(2)}_4 \!\!= 1.93\pm 0.1\ . \label{rockzeta}
\end{equation}
Two comments are in order. First, one should notice the
non-universality of the scaling exponents as compared with the
previous example (\ref{metzeta}). Second, the present surface does
{\em not} satisfy a $\theta\to -\theta$ symmetry. As the experiments
generating the surface were full dynamic, the fracture front
undergoes the well-known microbranching instability, resulting in
directed ``branch lines'' that break the $\theta\to -\theta$
symmetry \cite{02SCF}. Due to the lack of symmetry the amplitudes of
the coefficient $a_m$ can take any phase, not constrained to 0 or
$\pi$ as required by the $\theta\to -\theta$ symmetry. The lack of
symmetry is clearly obvious in the reconstruction of the structure
function from the irreducible representations. In Fig. \ref{r2} we
compare the experimental values of $S_2(\ell,\theta)$, for $\ell=2~
mm$, to the expansion

\begin{figure}
\epsfxsize=6 truecm \epsfbox{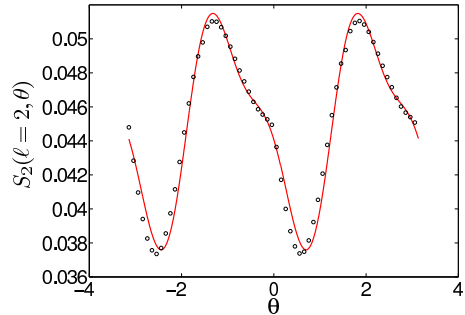}
 \caption{Comparison of the
experimental $S_2(\ell, \theta)$ (circles) and the SO(2) expansion
up to the third even irreducible representation (line) for the
artificial rock with $\ell=2 mm$.} \label{r2}
\end{figure}

\begin{eqnarray}
S_2( \ell, \theta)& \approx &0.025~ \ell^{0.86}  +0.0016 ~ \ell^{1.93} \cos(2\theta + 2.09) \nonumber\\
&+& 5.4\times 10^{-4} ~\ell ^{1.93} \cos(4\theta - 0.17) \ ,
\label{represent2}
\end{eqnarray}
The fit is satisfactory and the asymmetry in $\theta$ is obvious.

Taking the present two examples as representative, it appears that
the SO(2) decomposition extracts pure scaling behavior in each
sector, but that the scaling exponents are not universal, at least
in the two experiments discussed here. A possible explanation for
the discrepancy in the exponents is that the latter experiment was
fully dynamic whereas the former was quasi-static. Considering cuts
in $S_2(\ell,\theta)$ along the $\theta=0$ and $\theta=\pi$
directions, the present approach predicts a mixture of scaling
exponents rather than pure power laws, potentially leading to
spurious exponents.

Finally, one should point out that the SO(2) decomposition is not
expected to yield satisfactory results when the material itself is
strongly anisotropic. As an example we considered fracture surfaces
in wood. This is clearly an anisotropic medium due to the fiber
structure and indeed we found that along and across the fiber
directions the scaling behavior appears credible, whereas the SO(2)
decomposition failed altogether to reveal clean scaling properties
in any sector.

In summary, we propose that materials which can be fractured in an
isotropic fashion, i.e materials having an isotropic structure,
often have anisotropic fracture surfaces only because of the
breaking of isotropy by the initial conditions. In such cases it
appears useful to analyze the anisotropic contributions as
``corrections to scaling" beyond the isotropic sector, which is
always there, with a leading scaling exponent. The analysis
reveals non-universality in the scaling exponents, a finding that
calls for further future study and assessment, including the
interesting question of the possible existence of universality
classes. On the practical side, we have demonstrated that the full
information concerning the two dimensional structure function can
be efficiently parameterized by a few amplitudes and scaling
exponents. The reader should note that here we dealt only with
second order structure functions. In analogy to turbulence it may
be possible to decompose any higher order structure function into
SO(2) irreducible representations \cite{99ALP}. This may reveal
additional interesting scaling properties such as the phenomenon
of multiscaling discussed in Sec. \ref{multiscaling}. Finally, we
would like to emphasize the great interest in the proposed
isotropic fracture experiment and the measurement of the roughness
exponent in such an experiment. If indeed this scaling exponent
were identical to the exponent of the isotropic sector in a
standard experiment, this would significantly strengthen the
theoretical interest in the proposed approach.

\section{Summary}

The main thrust of this paper is that a careful study of the scaling
properties of the graphs representing fracture surfaces can lead to
a better understanding of the physics of fracture. We showed that in
1+1 dimensions fracture lines are multiscaling, a property that is
not reproduced in a number of traditional models of fracture.
Rather, one needs to consider possible failures of linear elasticity
in the vicinity of the crack tip, to form a typical scale ahead of
the crack tip characterizing the growth steps in the crack
propagation. Also in 2+1 dimensions we find, after decomposing the
statistical objects into their SO(2) irreducible representations, a
host of anomalous exponents in the various sectors that cannot be
understood without a careful rendering of the physics of fracture.
We thus propose that future research should focus on the mechanisms
for the failure of linear elasticity and how the physics discovered
 manifests itself in the scaling properties of the graphs of the fracture surfaces.
\label{Summary}

{\bf Acknowledgments} We are grateful to D. Bonamy, E. Bouchaud, J.
Fineberg and A. Sagy for sharing with us their valuable experimental
fracture surface data. We acknowledge useful discussions with D.
Bonamy, E. Brener, E. Bouchaud, S. Ciliberto, J. Fineberg, R.
Spatchek, J. R. Rice. E.B. is supported by the Horowitz Complexity
Science Foundation. This work was supported in part by the Israel
Science Foundation administered by the Israel Academy of Sciences.

\end{document}